\documentclass[12pt, draftclsnofoot, onecolumn]{IEEEtran}

\usepackage{amsmath,amsfonts,amsthm,amssymb}
\usepackage{bbm}
\usepackage{graphicx}
\usepackage{cite}
\usepackage{epstopdf}
\usepackage{float}
\usepackage{subfigure}
\usepackage{setspace}
\usepackage{stfloats}
\usepackage{url}

\begin{document}

\title{Towards Reliable UAV Swarm Communication in D2D-Enhanced Cellular Network}

\author{Yitao~Han,
	Liang~Liu,
        Lingjie~Duan,
        and~Rui~Zhang
\thanks{Y.~Han and L.~Duan are with the Engineering Systems and Design Pillar, Singapore University of Technology and Design (e-mail: yitao\_han@mymail.sutd.edu.sg, lingjie\_duan@sutd.edu.sg). Y.~Han is also with the Department of Electrical and Computer Engineering, National University of Singapore. L.~Duan is the corresponding author.}
\thanks{L.~Liu is with the Department of Electronic and Information Engineering, the Hong Kong Polytechnic University (e-mail: liang-eie.liu@polyu.edu.hk).}
\thanks{R.~Zhang is with the Department of Electrical and Computer Engineering, National University of Singapore (e-mail: elezhang@nus.edu.sg).}
}

\maketitle

\vspace{-3.8em}

\begin{abstract}
\begin{spacing}{1.4}

In the existing cellular networks, it remains a challenging problem to communicate with and control an unmanned aerial vehicle (UAV) swarm with both high reliability and low latency. Due to the UAV swarm's high working altitude and strong ground-to-air channels, it is generally exposed to multiple ground base stations (GBSs), while the GBSs that are serving ground users (occupied GBSs) can generate strong interference to the UAV swarm. To tackle this issue, we propose a novel two-phase transmission protocol by exploiting cellular plus device-to-device (D2D) communication for the UAV swarm. In Phase I, one swarm head is chosen for ground-to-air channel estimation, and all the GBSs that are not serving ground users (available GBSs) transmit a common control message to the UAV swarm simultaneously, using the same cellular frequency band. Both the swarm head and other swarm members can utilize the high power gain from multiple available GBSs' transmission, to combat the strong interference from occupied GBSs, while some UAVs may fail to decode the message due to uncorrelated ground-to-air channels. In Phase II, all the UAVs that have decoded the common control message in Phase I further relay it to the other UAVs in the swarm via D2D communication, by exploiting the less interfered D2D frequency band and the proximity among UAVs. In this paper, we aim to characterize the reliability performance of the above two-phase protocol, i.e., the expected percentage of UAVs in the swarm that can decode the common control message, which is a non-trivial problem due to the complex system setup and the intricate coupling between the two transmission phases. Nevertheless, we manage to obtain an approximated expression of the reliability performance of interest, under reasonable assumptions and with the aid of the Pearson distributions. Numerical results validate the accuracy of our analytical results and show the effectiveness of our protocol over other benchmark protocols. We also study the effect of key system parameters on the reliability performance, to reveal useful insights on the practical design of cellular-connected UAV swarm communication.

\end{spacing}
\end{abstract}

\vspace{-0.5em}


\section{Introduction}

The rapid development of unmanned aerial vehicle (UAV) has attracted a great deal of attention from both academia and industry in recent years. UAV possesses unique advantages comparing with other traditional terrestrial or aerial platforms, such as three-dimensional (3D) mobility, on-demand deployment, low cost, favourable air-to-ground channel, and so on \cite{zeng2016wireless}. As such, UAV is highly suitable for realizing various applications, e.g., aerial imagining, item delivery, traffic management, and serving as communication platform \cite{zeng2016wireless}. For the vast applications to turn into reality, it is crucial to ensure high-performance wireless communication with UAVs, for enabling their real-time command and control (C\&C) for safe operation as well as throughput-demanding payload communication with their associated ground pilots/users \cite{3GPP}. However, the existing UAV communication mainly relies on unlicensed spectrum, which is unreliable, limited in data rate, and only operable within visual/radio line-of-sight (LoS) range. While integrating UAVs with different applications into the existing (4G) and future (5G) cellular network can be a promising solution to the issues above \cite{zeng2019cellular,lin2018sky}, since it can offer a tremendous performance gain over the existing UAV communication, in terms of reliability, throughput, and operation range. Current research on cellular-connected UAV communication mainly focuses on how to improve the throughput of payload communication \cite{wu2018common,mei2019uplink,liu2018multi}, yet reliability and latency related issues in C\&C communication have drawn limited attention so far, albeit a handful of works on addressing them appeared recently \cite{kerczewski2012control,azari2018ultra,she2019ultra}.

Meanwhile, the capability of one single UAV is restricted due to its limited size and processing power, while UAV swarm can open up new opportunities, since multiple cooperative UAVs can accomplish complex missions, e.g., data collection, tracking-and-surveillance, and so on \cite{shakhatreh2019uav,bekmezci2013flying}. Therefore in this paper, we set the goal to communicate with and control a cellular-connected UAV swarm with both high reliability and low latency, i.e., all the UAVs in the swarm are to receive a common control message\footnote{Instead of sending dedicated control messages for individual UAVs, we consider the use of one common control message for the entire swarm. This is because the movements of the UAVs in the swarm are generally of high correlation, using one common control message can help reduce the total message size, as compared to dedicated control messages.} within the latency requirement in a reliable manner. However, there are challenges against the realization of the above goal. First, the UAV swarm's high working altitude leads to the strong ground-to-air channels from ground base stations (GBSs) to UAVs \cite{wu2018common,mei2019uplink,liu2018multi}, which renders the UAV swarm to suffer from the strong downlink interference from those occupied GBSs that are serving ground users at the same cellular frequency band.\footnote{Note that it is practically difficult to assign an exclusive frequency band for UAV swarm communication which is not being used by any GBS underneath, due to the high frequency reuse in the existing cellular network as well as the large number of GBSs involved because of the strong ground-to-air channels.} Second, the UAVs in the swarm are separated with certain distance for safety reason, which makes the ground-to-air channels uncorrelated in practice \cite{goldsmith2005wireless}. Therefore, some UAVs with low signal-to-interference-plus-noise-ratio (SINR) will fail to decode the common control message from cellular network. However, if we perform a paradigm shift, the above challenges can become opportunities. First, the UAV swarm is covered by multiple available GBSs that are not serving ground users, thus letting multiple available GBSs transmit the message to the UAV swarm simultaneously using the same cellular frequency band can provide a high power gain, to combat the strong interference from occupied GBSs. Second, despite the safety separation between UAVs, the UAV swarm is still dense, thus the UAVs inside the swarm can establish a reliable device-to-device (D2D) communication network for message sharing.

Motivated by the above, this paper proposes a novel two-phase transmission protocol that exploits cellular plus D2D communication, to achieve high reliability and low latency communication with the UAV swarm. In Phase I, all the available GBSs transmit a common control message to the UAV swarm simultaneously using the same cellular frequency band, to combat the strong interference from occupied GBSs. But the uncorrelated ground-to-air channels result in some UAVs failing to decode the message. In Phase II, all the UAVs that have decoded the common control message in Phase I further relay it to the other UAVs in the swarm via D2D communication, by exploiting the less interfered D2D frequency band \cite{u2018overview} and the proximity among UAVs.

\subsection{Prior Work}

Current studies on ultra-reliable and low-latency communication (URLLC) mainly focus on terrestrial communication. \cite{johansson2015radio} and \cite{bennis2018ultra} give an overview on URLLC and introduce the tools, methodologies, and possible solutions. \cite{sybis2016channel} compares the block error rate and computational complexity of different coding schemes for URLLC in 5G. \cite{vu2017ultra} investigates URLLC in mmWave-enabled massive multiple-input multiple-output networks. \cite{singh2017contention} studies uplink URLLC using contention-based approach, while \cite{ji2018ultra} considers URLLC in downlink transmission from a physical layer perspective. But as mentioned earlier, the channel of UAV communication is different from that of terrestrial communication \cite{wu2018common,mei2019uplink,liu2018multi}, and reliable UAV communication also faces unique challenges and opportunities, which need to be carefully addressed under new setups with new solutions.

On the other hand, there are existing works studying two-phase transmission protocol with the aid of D2D communication. \cite{hsu2017motivate} focuses on social grouping for content sharing from a network and game theory perspective, and \cite{liu2018d2d} focuses on optimizing the beamforming vector at base station to maximize the reliability performance. In our proposed two-phase transmission protocol, we assume that there are multiple GBSs simultaneously serving the UAV swarm in Phase I, and our aim is to obtain the analytical expression of our proposed protocol's reliability performance.

It is worth noting that there are only few papers considering reliability and latency related issues in cellular-connected UAV communication: \cite{azari2018ultra} considers the case with one single UAV, and \cite{she2019ultra} considers the case with non-cooperative UAVs, both without the presence of interference. While in this paper, we study how to communicate with and control a cellular-connected UAV swarm with both high reliability and low latency, subjected to the strong downlink interference from cellular network.

\subsection{Main Contributions}

The main contributions of this paper are summarized as follows. First, to our best knowledge, this is the first paper studying how to control a cellular-connected UAV swarm with both high reliability and low latency, and we propose a novel two-phase transmission protocol with the aid of D2D communication to achieve our goal.

Next, we manage to obtain an analytical expression of the reliability performance, i.e., the expected percentage of UAVs in the swarm that can decode the common control message with our proposed two-phase protocol, by averaging over the random locations of GBSs and UAVs as well as the channels in two transmission phases. However, due to the complex system setup and the intricate coupling between the two phases, standard stochastic geometry approach is not adequate for our analysis. Nevertheless, we show that under reasonable assumptions, it is sufficient to characterize the reliability performance of Phase I and Phase II separately. The Pearson distributions are further used to obtain an accurate and closed-form approximation.

Finally, we conduct extensive simulations to show the accuracy of our analytical results and the effectiveness of our protocol over other benchmark protocols. We demonstrate the effect of the swarm head, by extending to the case with multiple rounds of D2D communication. We also study the effect of key system parameters to reveal useful insights on the practical system design, e.g., the reliability performance is shown to decrease as the radius of the UAV swarm increases, while there exists a trade-off in deciding the transmission time of Phase I and that of Phase II, and we need to balance between them for optimizing the reliability performance.

\subsection{Organization}

The rest of this paper is organized as follows. In Section II, we present the system model. In Section III, we introduce the two-phase transmission protocol in detail. In Sections IV, V and VI, we derive an analytical expression of our proposed protocol's reliability performance. In Section VII, we provide numerical results to validate the accuracy of our analytical results, demonstrate the effectiveness of our proposed protocol, and reveal useful insights on the practical system design. In Section VIII, we conclude this paper.

\section{System Model}

\begin{figure}[!t]
\centering
\includegraphics[width=0.55\textwidth]{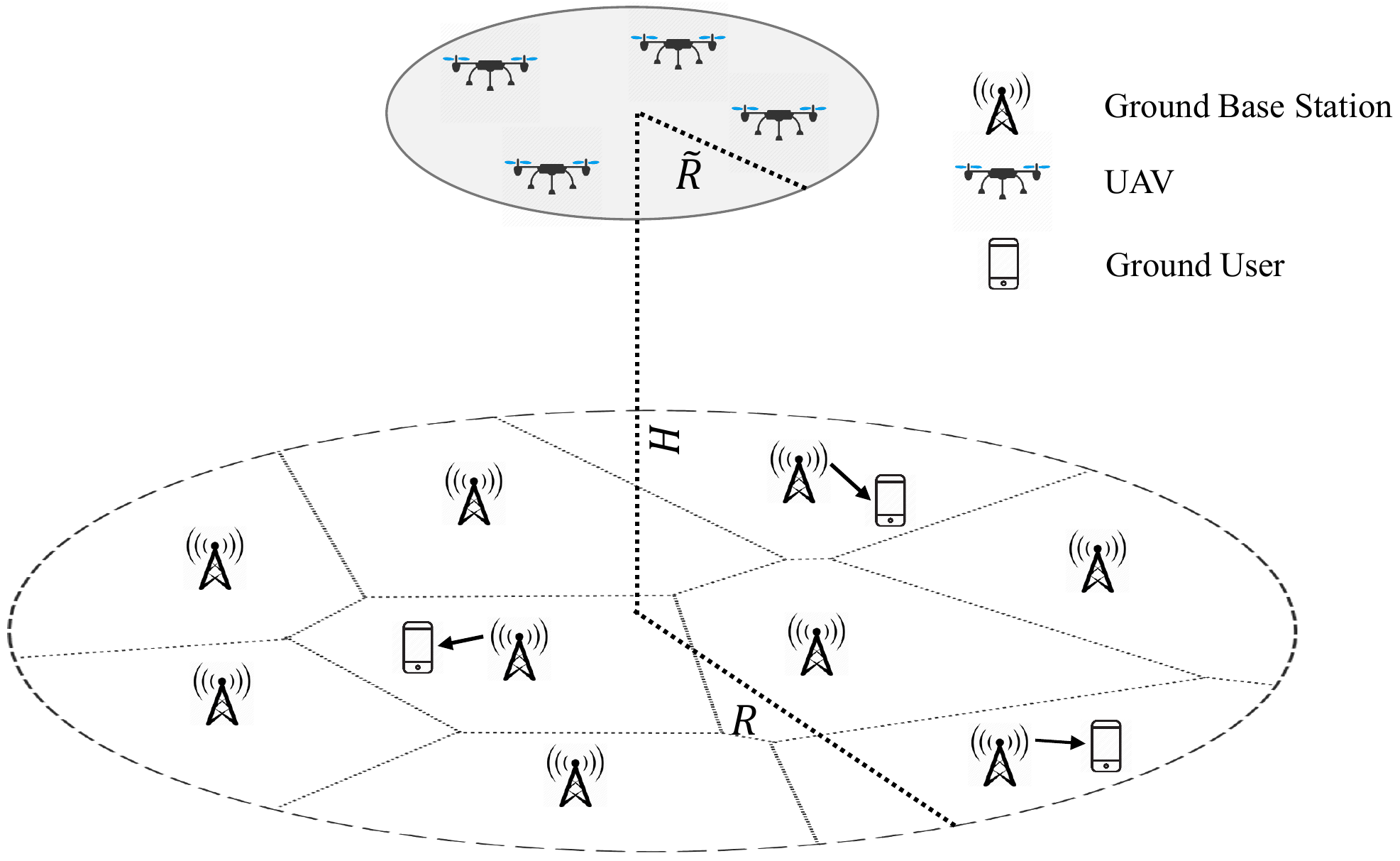}
\caption{Illustration of a cellular-connected UAV swarm (whose size is exaggerated comparing with the size of its signal coverage region).}
\label{sm}
\end{figure}

As illustrated in Fig.~\ref{sm}, we consider the downlink communication in a cellular network from a set of GBSs to a UAV swarm. For the purpose of exposition, it is assumed that each UAV is equipped with a single antenna, while each GBS employs an antenna array with fixed directional gain pattern \cite{3GPP}. The swarm includes $N$ UAVs, denoted by $\mathcal{N} = \{1,\cdots,N\}$. We assume that these UAVs are located inside a circular horizontal disk with radius $\tilde{R}$, altitude $H$,\footnote{In practice, the UAVs inside a swarm are more likely to be at similar altitudes, rather than the same altitude. Our analysis can also be extended to a 3D UAV swarm.} and center $\tilde{\mathbf{u}}\in\mathbb{R}^{3\times1}$, following a ``hard-core'' point process with a minimal separation distance $d_{\min}$ between any two UAVs to avoid collision \cite{haenggi2011mean}. The 3D location of UAV $n$ is denoted by $\tilde{\mathbf{u}}_n\in\mathbb{R}^{3\times1}$, $n\in\mathcal{N}$, and $\tilde{\mathbf{U}}=[\tilde{\mathbf{u}}_1,\cdots,\tilde{\mathbf{u}}_N]\in\mathbb{R}^{3\times N}$ is defined as the matrix containing all the locations of UAVs. Due to channel attenuation, the UAV swarm can only be covered by the GBSs inside a circular ground disk with radius ${R}$, and the center of the ground disk is exactly below the center of the UAV swarm $\tilde{\mathbf{u}}$. We assume that there are $M$ GBSs located inside this circular ground disk (coverage region), denoted by $\mathcal{M}=\{1,\cdots,M\}$, with their locations following a 2D binomial point process (BPP) independent of the point process for UAVs' locations. The 3D location of GBS $m$ is denoted by ${\mathbf{u}}_m\in\mathbb{R}^{3\times1}$, $m\in \mathcal{M}$, and $\mathbf{U}=[\mathbf{u}_1,\cdots,\mathbf{u}_M]\in\mathbb{R}^{3\times M}$ is defined as the matrix containing all the locations of GBSs. It is worth noting that since the UAV swarm is mobile, the locations of the disks containing the UAVs and their covered GBSs are moving in general, therefore the number of GBSs $M$ and even the number of UAVs in the swarm $N$ can change over time. However, in this paper we mainly consider a snap of the time, where the locations of the disks, also $M$ and $N$, are fixed, while the GBSs and UAVs are randomly located inside these disks, respectively. The results can be applied to any snap during the UAV swarm's flight.

The UAV swarm is allocated to one particular cellular frequency band for the downlink communication. $M$ GBSs are classified into two categories: $M_0<M$ available GBSs that are not serving ground users at this band, and $M_1=M-M_0$ occupied GBSs that are serving ground users at this band, respectively. As a result, the UAV swarm can be served by available GBSs subjected to the interference from occupied GBSs. We assume that the sets of available GBSs and occupied GBSs are independently generated, so that available GBSs' and occupied GBSs' locations follow independent BPPs, respectively. For simplicity, we label GBSs $1,\cdots,M_0$ as available GBSs, denoted by $\mathcal{M}_0=\{1,\cdots,M_0\}$, and GBSs $M_0+1,\cdots,M$ as occupied GBSs, denoted by $\mathcal{M}_1=\{M_0+1,\cdots,M\}$.

In this paper, we set the goal that all the UAVs inside the swarm are to receive a common control message of $D$ bits from the cellular network within an end-to-end latency of $\tau$ seconds (s) in a reliable manner. The common control message is intended to provide the UAV swarm with flight guidance and cooperation instruction. Due to the stringent latency requirement, $\tau$ is in general less than the channel coherence time. As a result, the ground-to-air channels are assumed to stay constant over each time slot of $\tau$ s. Note that although LoS path exists due to the UAV swarm's high working altitude, there are also scattered paths introduced by ground obstacles and other UAVs in the swarm. Therefore, a practical Rician fading channel model is considered between each GBS and each UAV for the downlink cellular communication. The equivalent complex baseband channel from GBS $m$ to UAV $n$ is denoted by
\begin{equation}
\begin{aligned}
h_{n,m}& = \sqrt{\frac{\beta}{d_{n,m}^{\alpha}}}h_{n,m}^{\text{Rician}}\\
& = \sqrt{\frac{\beta}{d_{n,m}^{\alpha}}}\bigg(\sqrt{\frac{\kappa}{\kappa+1}}h_{n,m}^{\text{LoS}}+\sqrt{\frac{1}{\kappa+1}}h_{n,m}^{\text{Rayleigh}}\bigg), \ m\in\mathcal{M}, n\in\mathcal{N},
\label{ii1}
\end{aligned}
\end{equation}where $\beta$ is the channel power gain at the reference distance $d_{\text{ref}}=1$ meter (m), $d_{n,m} = \|\tilde{\bold{u}}_n-\bold{u}_m\|_2$ is the Euclidean distance between GBS $m$ and UAV $n$, $\alpha$ is the path loss exponent of the ground-to-air channels, $h_{n,m}^{\text{Rician}}$ is the normalized Rician fading channel, $\kappa$ is the Rician factor specifying the power ratio between LoS and Rayleigh fading components, $h_{n,m}^{\text{LoS}}$ with $|h_{n,m}^{\text{LoS}}|=1$ is the normalized LoS component, and $h_{n,m}^{\text{Rayleigh}}\sim\mathcal{CN}(0,1)$ is the normalized Rayleigh fading component. We further define $\mathbf{H}\in\mathbb{C}^{N\times M}$ as the matrix containing all the fading channels from GBSs to UAVs. Note that the channels from different GBSs to different UAVs are uncorrelated, because in practice the distance between GBSs and the minimal separation between UAVs are much larger than half of the carrier wavelength.

\section{Two-phase Transmission Protocol for UAV Swarm}

To communicate with and control the UAV swarm in a reliable manner, it is desirable that on average (over the random locations of GBSs and UAVs, and their channels in the two transmission phases) a large portion of UAVs in the swarm, e.g., $99\%$ or even higher, can decode the common control message within the latency requirement. However, in practice, it is challenging to achieve the above goal because of the strong downlink interference from co-channel occupied GBSs and the uncorrelated ground-to-air channels.

\begin{figure}[t]
  \centering
  \subfigure[Phase I]{
    \includegraphics[width=0.45\textwidth]{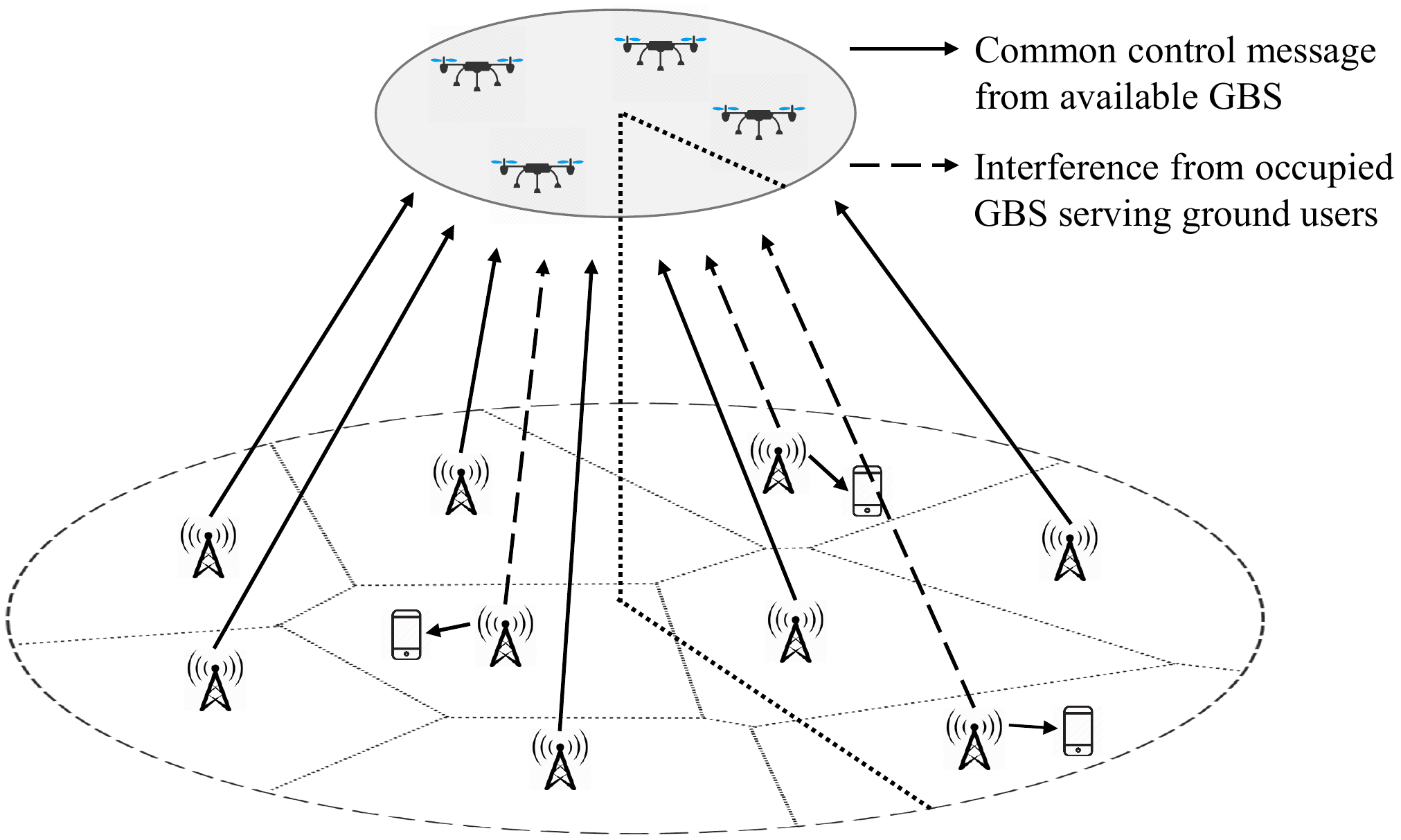}}
  \hspace{0.1in}
  \subfigure[Phase II]{
    \includegraphics[width=0.45\textwidth]{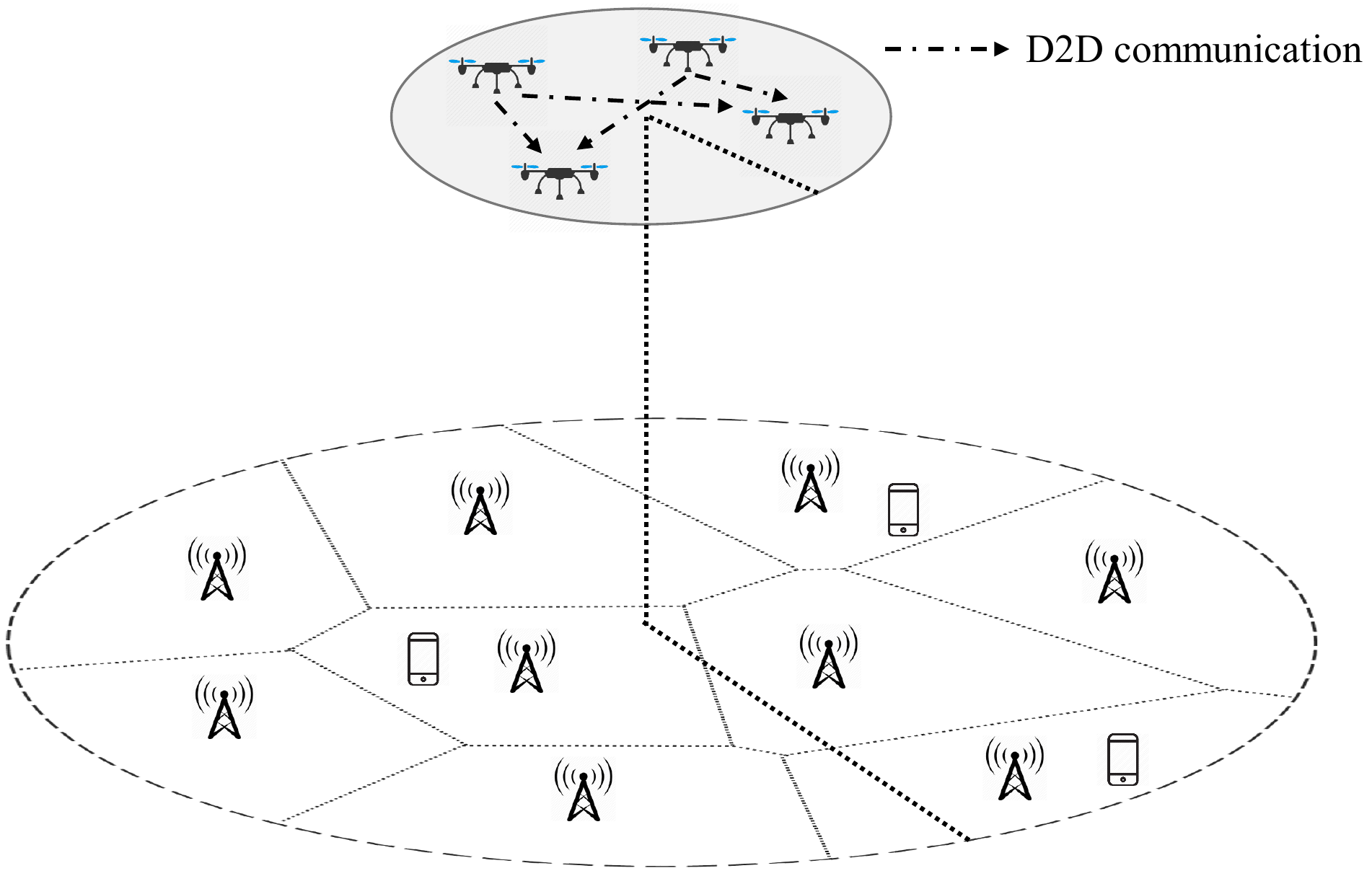}}
  \caption{Two-phase transmission protocol in D2D-enhanced cellular network: in Phase I, available GBSs transmit the common control message to the UAV swarm, and in Phase II, the successfully decoded UAVs relay the message to those that failed to decode it in Phase I via D2D communication.}
  \label{smp}
\end{figure}

To overcome the above issues, we propose a novel two-phase transmission protocol, as shown in Fig.~\ref{smp}. Specifically, in Phase I with duration $\tau^{(\text{I})}<\tau$, all the available GBSs inside the UAV swarm's coverage region transmit the common control message to the UAV swarm, then in Phase II with duration $\tau^{(\text{II})}=\tau-\tau^{(\text{I})}$, the UAVs that have decoded the message in Phase I further help relay it to the other UAVs in the swarm via D2D communication. We assume that the short-range D2D communication utilizes a frequency band that is not used by the cellular network, to avoid the strong interference from occupied GBSs as well as causing interference to the cellular communication with ground users \cite{u2018overview}. This provides a new diversity gain, since the UAVs that failed to decode the message in Phase I can leverage the less interfered D2D communication to decode it in Phase II.

In the following, we introduce the two-phase transmission protocol in detail.

\subsection{Phase I: Transmission from GBSs to UAV Swarm}

As shown in Fig.~\ref{smp} (a), in Phase I, all the available GBSs transmit the common control message to the UAV swarm simultaneously using the same cellular frequency band,\footnote{It is assumed that all the available GBSs can receive the common control message before Phase I, and they are synchronized for simultaneous transmission in Phase I through high-speed backbone network.} to combat the strong interference from occupied GBSs. In our considered protocol, one swarm head is chosen for channel estimation with available GBSs, labelled as UAV $1$. It is assumed that channel reciprocity holds between the uplink and downlink communications, such that all the available GBSs can know the channel state information (CSI) of the swarm head simply by letting it send a pilot in the uplink. Based on the estimated CSI, the transmitted signal of available GBS $m_0$ is given by
\begin{equation}
\begin{aligned}
x_{m_0}^{(\text{I})} = \sqrt{P}\omega_{m_0}s,\ \ \ m_0\in\mathcal{M}_0,
\label{iii2}
\end{aligned}
\end{equation}where $P$ is the identical transmit power of GBSs, $s$ is the common control message, with $\mathbb{E}\{|s|^2\}=1$, and $\omega_{m_0}$ is the equal-gain combining weight for the swarm head at available GBS $m_0$, i.e.,
\begin{equation}
\begin{aligned}
\omega_{m_0} = \frac{h^*_{1,{m_0}}}{\big|h_{1,{m_0}}\big|},
\label{iii3}
\end{aligned}
\end{equation}
where $(\cdot)^*$ is the conjugate operation.

Note that occupied GBSs serve ground users under the same cellular frequency band. The transmitted signal of occupied GBS $m_1$ is given by
\begin{equation}
\begin{aligned}
x_{m_1}^{(\text{I})} = \sqrt{P}s_{m_1},\ \ \ m_1\in\mathcal{M}_1,
\label{iii4}
\end{aligned}
\end{equation}where $s_{m_1}$ is occupied GBS $m_1$'s transmitted message intended for its served ground users, with $\mathbb{E}\{|s_{m_1}|^2\}=1$, which is assumed to be independent over $m_1$'s as well as $s$.

Combining \eqref{iii2} and \eqref{iii4}, the received signal at UAV $n$ in Phase I is
\begin{equation}
\begin{aligned}
y_{n}^{(\text{I})} &= \sum_{m_0\in\mathcal{M}_0}h_{n,m_0}x_{m_0}^{(\text{I})} + \sum_{m_1\in\mathcal{M}_1}h_{n,m_1}x_{m_1}^{(\text{I})}+z_{n}^{(\text{I})}\\
& =\sqrt{P}\sum_{m_0\in\mathcal{M}_0}\frac{h_{n,m_0}h^*_{1,{m_0}}}{\big|h_{1,{m_0}}\big|}s+\sqrt{P}\sum_{m_1\in\mathcal{M}_1}h_{n,m_1}s_{m_1}+z_{n}^{(\text{I})},\ \ \ n\in\mathcal{N},
\label{iii5}
\end{aligned}
\end{equation}where $z_{n}^{(\text{I})}\sim\mathcal{CN}(0,\sigma^{2})$ is the additive white Gaussian noise (AWGN) at UAV $n$.

As a result, the SINR at the swarm head in Phase I is
\begin{equation}
\begin{aligned}
{\tt{SINR}_{1}^{(\text{I})}}  = \frac{P\big|\sum_{m_0\in\mathcal{M}_0}|{h}_{1,m_0}|\big|^2}{P\sum_{m_1\in\mathcal{M}_1}|h_{1,m_1}|^2+\sigma^{2}}.
\label{iii6}
\end{aligned}
\end{equation}
On the other hand, the SINRs at the other swarm members in Phase I are
\begin{equation}
\begin{aligned}
{\tt{SINR}_{n}^{(\text{I})}} = \frac{P\Big|\sum_{m_0\in\mathcal{M}_0}\frac{h_{n,m_0}h^*_{1,{m_0}}}{|h_{1,{m_0}}|}\Big|^2}{P\sum_{m_1\in\mathcal{M}_1}|h_{n,m_1}|^2+\sigma^{2}},\ \ \ n\in\mathcal{N}\backslash\{1\}.
\label{iii7}
\end{aligned}
\end{equation}Note that due to dedicatedly designed beamforming, in general the swarm head can achieve a sufficiently high SINR in Phase I, as in \eqref{iii6}. While the uncorrelated ground-to-air channels make the SINRs in Phase I uncorrelated for different swarm members, thus it is expected that some swarm members can also achieve a reasonably high SINR in Phase I.

Recall that in Phase I with duration $\tau^{(\text{I})}$ s, the UAVs aim to decode the common control message of $D$ bits, i.e., the following constraints need to be satisfied
\begin{equation}
\begin{aligned}
\tau^{(\text{I})}B\log_2(1+\rho\times{\tt{SINR}_{n}^{(\text{I})}})\geq D,\ \ \ n\in\mathcal{N}. \nonumber
\end{aligned}
\end{equation}
Thus, the SINR requirement at each UAV in Phase I can be expressed as
\begin{equation}
\begin{aligned}
\theta^{(\text{I})} = \frac{2^{\frac{D}{\tau^{(\text{I})}B}}-1}{\rho},
\label{iii8}
\end{aligned}
\end{equation}where $B$ is the transmission bandwidth in Phase I, and $\rho\in(0,1)$ is the SINR gap from the channel capacity due to the usage of practical modulation and/or coding \cite{goldsmith2005wireless}. If ${\tt{SINR}_{n}^{(\text{I})}}\geq\theta^{(\text{I})}$, then UAV $n$ can decode the message in Phase I. We define
\begin{equation}
\begin{aligned}
\mathbf{\Theta}^{(\text{I})} = \{n:{\tt{SINR}_{n}^{(\text{I})}}\geq\theta^{(\text{I})},n=1,\cdots,N\}
\label{iii9}
\end{aligned}
\end{equation}as the set of UAVs that can decode the common control message in Phase I.

\subsection{Phase II: D2D Communication among UAVs}

As shown in Fig.~\ref{smp} (b), in Phase II, we can leverage the UAVs that have decoded the common control message in Phase I, i.e., $k\in\mathbf{\Theta}^{(\text{I})}$ with $|\mathbf{\Theta}^{(\text{I})}|=K$, to help relay it to the other UAVs in the swarm at a D2D frequency band that is not utilized by the cellular network.

Different from the Rician fading channels from GBSs to UAVs in Phase I, we assume a Rayleigh fading channel model for the D2D communication among UAVs in Phase II, since the signal from one UAV can be blocked or scattered by the other UAVs locally in the swarm. Thus, the equivalent complex baseband channel from transmitting UAV $k$ to receiving UAV $n$ is denoted by
\begin{equation}
\begin{aligned}
g_{n,k}=\sqrt{\frac{\tilde{\beta}}{\tilde{d}_{n,k}^{\tilde{\alpha}}}}g_{n,k}^{\text{Rayleigh}},\ \ \ k\in\mathbf{\Theta}^{(\text{I})},\ n\in\mathcal{N}\backslash\mathbf{\Theta}^{(\text{I})},
\label{iii10}
\end{aligned}
\end{equation}
where $\tilde{\beta}$ is the channel power gain of the D2D channels at the reference distance $\tilde{d}_{\text{ref}}=1$ m, $\tilde{d}_{n,k} = \|\tilde{\bold{u}}_n-\tilde{\bold{u}}_k\|_2$ is the Euclidean distance between transmitting UAV $k$ and receiving UAV $n$, $\tilde{\alpha}$ is the path loss exponent of the D2D channels in Phase II, and $g_{n,k}^{\text{Rayleigh}}\sim\mathcal{CN}(0,1)$ is the normalized Rayleigh fading channel. Further, we define $\mathbf{G}\in\mathbb{C}^{(N-K)\times K}$ as the matrix containing all the D2D channels in Phase II. Similar to the ground-to-air channels, we assume that the D2D channels between UAVs are uncorrelated, because the minimal separation distance between UAVs is much larger than half of the D2D communication wavelength.

Due to independent Rayleigh fading, all the UAVs that have decoded the common control message in Phase I, i.e., $k\in\mathbf{\Theta}^{(\text{I})}$, help relay it to the other UAVs in the swarm in Phase II for a higher average power gain, such that receiving UAVs can decode the message in Phase II with a higher probability. Since the channel estimation for D2D communication is difficult and time-consuming given the limited time duration $\tau^{(\text{II})}$, we assume that each UAV does not know the CSI to the other UAVs in Phase II in our considered protocol. Therefore, transmitting UAV $k$ that has decoded the message in Phase I simply transmits
\begin{equation}
\begin{aligned}
x_{k}^{(\text{II})} = \sqrt{\tilde{P}}\tilde{s},\ \ \ k\in\mathbf{\Theta}^{(\text{I})},
\label{iii11}
\end{aligned}
\end{equation}where $\tilde{P}$ is the identical transmit power of UAVs, and $\tilde{s}$ is the common control message with $\mathbb{E}\{|\tilde{s}|^2\}=1$, which carries the same information as $s$.\footnote{Phase I and Phase II use different frequency bands, in turn different modulation schemes in general, therefore $s$ and $\tilde{s}$ are independent even they carry the same information. For this reason, we assume that the UAVs cannot perform maximal ratio combining (MRC) to combine the received signals in Phase I and Phase II in practice.}

It is assumed that the UAVs have built-in clock and the above transmissions are synchronized. As a result, in Phase II, the received signal at receiving UAV $n$ that failed to decode the message in Phase I is given by
\begin{equation}
\begin{aligned}
y_{n}^{(\text{II})}& = \sum_{k\in\mathbf{\Theta}^{(\text{I})}}g_{n,k}x_{k}^{(\text{II})}+z_{n}^{(\text{II})}\\
& = \sqrt{\tilde{P}}\sum_{k\in\mathbf{\Theta}^{(\text{I})}}g_{n,k}\tilde{s}+z_{n}^{(\text{II})},\ \ \ n\in \mathcal{N}\backslash\mathbf{\Theta}^{(\text{I})},
\label{iii12}
\end{aligned}
\end{equation}where $z_{n}^{(\text{II})}\sim\mathcal{CN}(0,\tilde{\sigma}^{2})$ is the superposition of interference and AWGN at receiving UAV $n$ in Phase II. Note that the interference to the receiving UAVs in Phase II may come from other D2D networks operating at the same D2D frequency band \cite{u2018overview}, e.g., nearby UAV swarms, which is generally much smaller than the interference in Phase I from the occupied GBSs with much higher transmit power.

The corresponding SINR at receiving UAV $n$ in Phase II is
\begin{equation}
\begin{aligned}
{\tt{SINR}_{n}^{(\text{II})}} = \frac{\tilde{P}\big|\sum_{k\in\mathbf{\Theta}^{(\text{I})}}g_{n,k}\big|^2}{\tilde{\sigma}^{2}},\ \ \ n\in \mathcal{N}\backslash\mathbf{\Theta}^{(\text{I})}.
\label{iii13}
\end{aligned}
\end{equation}

Similar to \eqref{iii8}, the SINR requirement at each receiving UAV in Phase II can be expressed as
\begin{equation}
\begin{aligned}
\theta^{(\text{II})} = \frac{2^{\frac{D}{\tau^{(\text{II})}\tilde{B}}}-1}{\tilde{\rho}},
\label{iii14}
\end{aligned}
\end{equation}where $\tilde{B}$ is the transmission bandwidth in Phase II, $\tilde{\rho}$ is the SINR gap for the D2D communication. We can then define $\mathbf{\Theta}^{(\text{II})}$ as the set of UAVs that failed to decode the common control message in Phase I but can decode it in Phase II, i.e.,
\begin{equation}
\begin{aligned}
\mathbf{\Theta}^{(\text{II})} = \{n:{\tt{SINR}_{n}^{(\text{I})}}<\theta^{(\text{I})},{\tt{SINR}_{n}^{(\text{II})}}\geq\theta^{(\text{II})},n=1,\cdots,N\}.
\label{iii15}
\end{aligned}
\end{equation}

To summarize, for the UAVs that failed to decode the common control message in Phase I, Phase II provides another opportunity to decode it with a high reliability, thanks to the less interfered D2D channels and the proximity among UAVs.

\section{Performance Characterization}

In this section, we analytically characterize the expected percentage of UAVs in the swarm that can decode the common control message with our proposed two-phase transmission protocol, by averaging over the random locations of GBSs and UAVs, i.e., $\{{\bold{U}},\tilde{\bold{U}}\}$, as well as their fading channels in Phase I and Phase II, i.e., $\{\bold{H},\bold{G}\}$, which can be denoted by
\begin{equation}
\begin{aligned}
\eta& = \frac{1}{N}\mathbb{E}_{\{{\bold{U}},\tilde{\bold{U}}, \bold{H},\bold{G}\}}\big\{\big|\mathbf{\Theta}^{(\text{I})}\cup\mathbf{\Theta}^{(\text{II})}\big|\big\}\\
& = \frac{1}{N}\Big(\mathbb{E}_{\{{\bold{U}},\tilde{\bold{U}}, \bold{H}\}}\big\{\big|\mathbf{\Theta}^{(\text{I})}\big|\big\}+\mathbb{E}_{\{{\bold{U}},\tilde{\bold{U}},\bold{H},\bold{G}\}}\big\{\big|\mathbf{\Theta}^{(\text{II})}\big|\big\}\Big).
\label{iv16}
\end{aligned}
\end{equation}Clearly, the expected number of UAVs that can decode the message in Phase I, i.e., $\mathbb{E}\{|\mathbf{\Theta}^{(\text{I})}|\}$, and the expected number of UAVs that failed to decode the message in Phase I but can decode it in Phase II, i.e., $\mathbb{E}\{|\mathbf{\Theta}^{(\text{II})}|\}$, are coupled, which makes our analysis very difficult. In the following, we simplify \eqref{iv16} by tight approximations. Based on the key observation that the radius of the UAV swarm $\tilde{R}$ is in general much smaller than the radius of the coverage region ${R}$ as well as the altitude of the UAV swarm $H$, we have the following assumption.

\underline{\emph{Assumption 4.1}}: \
The Euclidean distances between GBS $m$ and all the $N$ UAVs in the swarm are assumed to be the same, which are equal to the Euclidean distance between GBS $m$ and the center of the UAV swarm, i.e.,
\begin{equation}
\begin{aligned}
{d}_{n,m}\approx {d}_m = \|\tilde{\bold{u}}-{\bold{u}}_m\|_2, \ \ m\in\mathcal{M},\ n\in\mathcal{N}.
\label{iv17}
\end{aligned}
\end{equation}

Given Assumption 4.1, $\mathbb{E}\{|\mathbf{\Theta}^{(\text{I})}|\}$ in \eqref{iv16} is only determined by the locations of GBSs, $\bold{U}$, and the fading channels in Phase I, $\bold{H}$, which will greatly ease our analysis of Phase I. Therefore, $\eta$ in \eqref{iv16} can be rewritten as
\begin{equation}
\begin{aligned}
\eta
&\approx\frac{1}{N}\Big(\mathbb{E}_{\{{\bold{U}}, \bold{H}\}}\big\{\big|\mathbf{\Theta}^{(\text{I})}\big|\big\}+\mathbb{E}_{\{{\bold{U}}, \tilde{\bold{U}}, \bold{H}, \bold{G}\}}\big\{\big|\mathbf{\Theta}^{(\text{II})}\big|\big\}\Big)\\
&\overset{\text{(a)}}{=}\frac{1}{N}\ \mathbb{E}_{\{\bold{d}, \bold{H}^{\text{Rician}}\}}\Big\{\big|\mathbf{\Theta}^{(\text{I})}\big|+\mathbb{E}_{\{\tilde{\bold{D}},\bold{G}^{\text{Rayleigh}}\}}\big\{\big|\mathbf{\Theta}^{(\text{II})}\big|\Big|\mathbf{\Theta}^{(\text{I})}\big\}\Big\}\\
&\overset{\text{(b)}}{=}\frac{1}{N}\ \mathbb{E}_{\{\bold{d}, \bold{H}^{\text{Rician}}\}}\Big\{\big|\mathbf{\Theta}^{(\text{I})}\big|+\mathbb{E}_{\{\tilde{\bold{D}},\bold{G}^{\text{Rayleigh}}\}}\big\{\big|\mathbf{\Theta}^{(\text{II})}\big|\Big|\big|\mathbf{\Theta}^{(\text{I})}\big|\big\}\Big\}.
\label{iv18}
\end{aligned}
\end{equation}In the above, we define ${\bold{d}}\in\mathbb{R}^{M\times1}$ as the vector containing all the distances between GBSs and the UAV swarm $d_m$'s, $\tilde{\bold{D}}\in\mathbb{R}^{(N-K)\times K}$ as the matrix containing all the distances between transmitting UAVs and receiving UAVs $\tilde{d}_{n,k}$'s, $\bold{H}^{\text{Rician}}\in\mathbb{C}^{N\times M}$ and $\bold{G}^{\text{Rayleigh}}\in\mathbb{C}^{(N-K)\times K}$ as the matrices containing all the small-scale fadings in Phase I, ${h}_{n,m}^{\text{Rician}}$'s, and Phase II, $g_{n,k}^{\text{Rayleigh}}$'s, respectively. The reasons for $\overset{\text{(a)}}{=}$ and $\overset{\text{(b)}}{=}$ to hold are given as follows.

$\overset{\text{(a)}}{=}$: In the channel from GBS $m$ to UAV $n$, i.e., ${h}_{n,m} = \sqrt{\frac{{\beta}}{{d}_{m}^{{\alpha}}}}{h}_{n,m}^{\text{Rician}}$, path loss $\sqrt{\frac{{\beta}}{{d}_{m}^{{\alpha}}}}$ and small-scale fading ${h}_{n,m}^{\text{Rician}}$ are independent. While the location of GBS $m$, $\bold{u}_m$, only determines the distance between GBS $m$ and the UAV swarm, ${d}_{m}$, which is already included in ${h}_{n,m}$. Therefore for $\mathbb{E}\{|\mathbf{\Theta}^{(\text{I})}|\}$, it is sufficient to average over path loss, $\bold{d}$, and small-scale fading, $\bold{H}^{\text{Rician}}$, which is equivalent to average over the locations of GBSs, ${\bold{U}}$, and the fading channels in Phase I, $\bold{H}$. Similarly for $\mathbb{E}\{|\mathbf{\Theta}^{(\text{II})}|\}$, given $\mathbf{\Theta}^{(\text{I})}$, it is sufficient to average over $\tilde{\bold{D}}$ and $\bold{G}^{\text{Rayleigh}}$, instead of $\tilde{\bold{U}}$ and $\bold{G}$.

$\overset{\text{(b)}}{=}$: The UAVs are randomly located inside the swarm, and all the successful decoded UAVs help transmit in Phase II with identical transmit power. Therefore, to obtain $\mathbb{E}\{|\mathbf{\Theta}^{(\text{II})}|\}$ by averaging over $\tilde{\bold{D}}$ and $\bold{G}^{\text{Rayleigh}}$, it is sufficient to know how many UAVs have decoded the message in Phase I, i.e., $|\mathbf{\Theta}^{(\text{I})}|$, rather than their exact indexes, i.e., $\mathbf{\Theta}^{(\text{I})}$.

The advantage of approximating \eqref{iv16} by \eqref{iv18} is the ``decouple'' of Phase I and Phase II. Specifically, to characterize $\eta$ in \eqref{iv18}, it is sufficient to derive the expected number of UAVs that can decode the message in Phase I, i.e., $\mathbb{E}_{\{\bold{d}, \bold{H}^{\text{Rician}}\}}\{|\mathbf{\Theta}^{(\text{I})}|\}$, and then the expected number of UAVs that can decode the message in Phase II given that $|\mathbf{\Theta}^{(\text{I})}|$ UAVs have decoded it in Phase I, i.e., $\mathbb{E}_{\{\tilde{\bold{D}}, \bold{G}^{\text{Rayleigh}}\}}\{|\mathbf{\Theta}^{(\text{II})}|\big||\mathbf{\Theta}^{(\text{I})}|\}$. In the following two sections, we characterize $\mathbb{E}_{\{\bold{d}, \bold{H}^{\text{Rician}}\}}\{|\mathbf{\Theta}^{(\text{I})}|\big\}$ and $\mathbb{E}_{\{\bold{d}, \bold{H}^{\text{Rician}}\}}\big\{\mathbb{E}_{\{\tilde{\bold{D}}, \bold{G}^{\text{Rayleigh}}\}}\{|\mathbf{\Theta}^{(\text{II})}|\big||\mathbf{\Theta}^{(\text{I})}|\}\big\}$, respectively.

\section{Analysis of Phase I}

In this section, we aim to characterize the expected number of UAVs that can decode the common control message in Phase I, $\mathbb{E}_{\{\bold{d}, \bold{H}^{\text{Rician}}\}}\{|\mathbf{\Theta}^{(\text{I})}|\}$. Thanks to the linearity of expectation, $\mathbb{E}_{\{\bold{d}, \bold{H}^{\text{Rician}}\}}\{|\mathbf{\Theta}^{(\text{I})}|\}$ can be expressed as the summation of the probability that each UAV can decode the message in Phase I, i.e.,
\begin{equation}
\begin{aligned}
\mathbb{E}_{\{\bold{d}, \bold{H}^{\text{Rician}}\}}\big\{\big|\mathbf{\Theta}^{(\text{I})}\big|\big\} =
\sum_{n\in\mathcal{N}}\mathbb{P}\big\{{\tt{SINR}_{n}^{(\text{I})}}\geq\theta^{(\text{I})}\big\}.\nonumber
\end{aligned}
\end{equation}Recall that in Phase I, the swarm head enjoys equal-gain combining beamforming gain from all the available GBSs, while the swarm members opportunistically decode the message from available GBSs. As a result, the decoding probability of the swarm head and that of the swarm members are different, and thus need to be analyzed separately.

\subsection{Decoding Probability of Swarm Head}

For simplicity, we assume AWGN is omitted in the analysis, since it is much smaller than the downlink interference in Phase I \cite{heath2013modeling}. As such, the SINR at the swarm head in Phase I given in \eqref{iii6} can be approximated by
\begin{equation}
\begin{aligned}
{\tt{SINR}_{1}^{(\text{I})}} \approx \frac{P\big|\sum_{m_0\in\mathcal{M}_0}|{h}_{1,m_0}|\big|^2}{P\sum_{m_1\in\mathcal{M}_1}|{h}_{1,m_1}|^2}
= \frac{\Big|\sum_{m_0\in\mathcal{M}_0}{d}_{m_0}^{\ \frac{-{\alpha}}{2}}|{h}_{1,m_0}^{\text{Rician}}|\Big|^2}{\sum_{m_1\in\mathcal{M}_1}{d}_{m_1}^{-{\alpha}}|{h}_{1,m_1}^{\text{Rician}}|^2}.
\label{iv19}
\end{aligned}
\end{equation}

We first analyze the numerator of \eqref{iv19}, i.e., $\overline{X} \triangleq \big|\sum_{m_0\in\mathcal{M}_0}{d}_{m_0}^{\ \frac{-{\alpha}}{2}}|{h}_{1,m_0}^{\text{Rician}}|\big|^2$. For convenience, we also define $\overline{X}_{int} \triangleq \sum_{m_0\in\mathcal{M}_0}{d}_{m_0}^{\ \frac{-{\alpha}}{2}}|{h}_{1,m_0}^{\text{Rician}}|$, i.e., $\overline{X}=\big|\overline{X}_{int}\big|^2$. Note that the exact distribution of $\overline{X}_{int}$ can be obtained by standard stochastic geometry approach, i.e., leveraging the Laplace transformation of ${d}_{m_0}^{\ \frac{-{\alpha}}{2}}|{h}_{1,m_0}^{\text{Rician}}|$'s. However, the result cannot be obtained in closed-form, which makes our further analysis difficult. Another commonly used approach is Gaussian-Chebyshev based approximation \cite{ding2013use}, which can provide a good approximation, but results in a long series and thus is too complicated for further analysis. In contrast, \cite{schmeiser1977methods} shows that the Pearson distributions (a family of continuous probability functions) can provide accurate approximations to certain empirical distribution functions. Motivated by this, we adopt the Pearson type III distribution, i.e., the Gamma distribution \cite{heath2013modeling}, to approximate the distribution of $\overline{X}_{int}$, where the parameters of the Gamma distribution can be obtained by matching $\overline{X}_{int}$'s first and second moments, also known as moment matching.

\underline{\emph{Lemma 5.1}}: \ The probability density function (PDF) of $\overline{X}_{int} = \sum_{m_0\in\mathcal{M}_0}{d}_{m_0}^{\ \frac{-{\alpha}}{2}}|{h}_{1,m_0}^{\text{Rician}}|$ can be approximated by
\begin{equation}
\begin{aligned}
f_{\overline{X}_{int}}(\overline{x}_{int}) = \frac{b_{\overline{X}}^{(a_{\overline{X}})}}{\Gamma(a_{\overline{X}})}\overline{x}_{int}^{(a_{\overline{X}}-1)}e^{-b_{\overline{X}}\overline{x}_{int}}, \ \ \ \overline{x}_{int}\geq 0, \nonumber
\end{aligned}
\end{equation}where $a_{\overline{X}}$ and $b_{\overline{X}}$ are the parameters of the Gamma distribution. Therefore, the approximated cumulative density function (CDF) of the numerator of \eqref{iv19}, $\overline{X} = \big|\overline{X}_{int}\big|^2$, is
\begin{equation}
\begin{aligned}
F_{\overline{X}}(\overline{x}) = \frac{\gamma\big(a_{\overline{X}},b_{\overline{X}}\sqrt{\overline{x}}\big)}{\Gamma(a_{\overline{X}})}, \ \ \ \overline{x}\geq 0,
\label{iv20}
\end{aligned}
\end{equation}where $\gamma(a,z)=\int_0^zt^{a-1}e^{-t}\,dt$ is the lower incomplete Gamma function [27, Eq. 8.350.1].
\begin{IEEEproof}
See Appendix A.
\end{IEEEproof}

Next, we address similarly the denominator of \eqref{iv19}, i.e., $\underline{X} \triangleq \sum_{m_1\in\mathcal{M}_1}{d}_{m_1}^{-{\alpha}}|{h}_{1,m_1}^{\text{Rician}}|^2$.

\underline{\emph{Lemma 5.2}}: \ The CDF of $\underline{X} = \sum_{m_1\in\mathcal{M}_1}{d}_{m_1}^{-{\alpha}}|{h}_{1,m_1}^{\text{Rician}}|^2$ can be approximated by
\begin{equation}
\begin{aligned}
F_{\underline{X}}(\underline{x}) = \frac{\gamma\big(a_{\underline{X}},b_{\underline{X}}\underline{x}\big)}{\Gamma(a_{\underline{X}})}, \ \ \ \underline{x}\geq 0.
\label{iv21}
\end{aligned}
\end{equation}
\begin{IEEEproof}
See Appendix B.
\end{IEEEproof}

Given Lemmas 5.1 and 5.2, we have the following proposition.

\underline{\emph{Proposition 5.1}}: \ The probability that the swarm head can decode the common control message in Phase I is
\begin{equation}
\begin{aligned}
& \mathcal{P}^{(\text{I-H})}(\theta^{(\text{I})}) = \mathbb{P}\bigg\{{\tt{SINR}_{1}^{(\text{I})}}=\frac{\overline{X}}{\underline{X}}\geq\theta^{(\text{I})}\bigg\} = 1 - F_{\frac{\overline{X}}{\underline{X}}}(\theta^{(\text{I})})\\
& = 1-\int_{0}^{\infty}\bigg(\int_{0}^{\theta^{(\text{I})}\underline{x}}f_{\overline{X}}(\overline{x})\,d\overline{x}\bigg)f_{\underline{X}}(\underline{x})\,d\underline{x}\\
& = 1-\frac{b_{\underline{X}}^{(a_{\underline{X}})}}{\Gamma(a_{\underline{X}})\Gamma(a_{\overline{X}})}\int_{0}^{\infty}\gamma\Big(a_{\overline{X}},b_{\overline{X}}\sqrt{\theta^{(\text{I})}}\sqrt{\overline{x}}\Big)\underline{x}^{(a_{\underline{X}}-1)}e^{-b_{\underline{X}}\underline{x}}\,d\underline{x}\\
& = 1-\frac{\big(\frac{b_{\overline{x}}^2\theta^{(\text{I})}}{b_{\underline{x}}}\big)^{\frac{1}{2}a_{\overline{x}}}}{\Gamma(a_{\underline{X}})\Gamma(a_{\overline{X}})}\bigg(-\frac{b_{\overline{x}}\sqrt{\theta^{(\text{I})}}\Gamma(\frac{1}{2}+a_{\underline{x}}+\frac{a_{\overline{x}}}{2})}{(1+a_{\overline{x}})\sqrt{b_{\underline{x}}}}{_2F_2}\Big(\frac{1}{2}+a_{\underline{x}}+\frac{a_{\overline{x}}}{2},\frac{1}{2}+\frac{a_{\overline{x}}}{2};\frac{3}{2},\frac{3}{2}+\frac{a_{\overline{x}}}{2};\frac{b_{\overline{x}}^2\theta^{(\text{I})}}{4b_{\underline{x}}}\Big)\\
&\ \ \ \ \ \ \ \ \ \ \ \ \ \ \ \ \ \ \ \ \ \ \ +\frac{\Gamma(a_{\underline{x}}+\frac{a_{\overline{x}}}{2})}{a_{\overline{x}}}{_2F_2}\Big(\frac{a_{\overline{x}}}{2},a_{\underline{x}}+\frac{a_{\overline{x}}}{2};\frac{1}{2},\frac{1}{2}+\frac{a_{\overline{x}}}{2};\frac{b_{\overline{x}}^2\theta^{(\text{I})}}{4b_{\underline{x}}}\Big)\bigg),
\label{iv22}
\end{aligned}
\end{equation}where ${_2F_2}(a,b;c,d;z)=\sum_{n=0}^{\infty}\frac{(a)_n(b)_n}{(c)_n(d)_n}\frac{z^n}{n!}$ is the generalized hypergeometric series [27, Eq. 9.14.1], with $(a)_0=1$ and $(a)_n=a(a+1)(a+2)\cdots(a+n-1)$ for $n\geq1$.
\begin{IEEEproof}
See Appendix C.
\end{IEEEproof}

\subsection{Decoding Probability of Swarm Members}

By ignoring the AWGN, the SINR at each swarm member in Phase I given in \eqref{iii7} can be approximately rewritten as
\begin{equation}
\begin{aligned}
{\tt{SINR}_{n}^{(\text{I})}} \approx \frac{P\big|\sum_{m_0\in\mathcal{M}_0}\frac{h_{n,m_0}h^*_{1,{m_0}}}{|h_{1,{m_0}}|}\big|^2}{P\sum_{m_1\in\mathcal{M}_1}|h_{n,m_1}|^2}
 = \frac{\Big|\sum_{m_0\in\mathcal{M}_0}\frac{{d}_{m_0}^{\ \frac{-{\alpha}}{2}}{h}_{n,m_0}^{\text{Rician}}h^*_{1,{m_0}}}{|h_{1,{m_0}}|}\Big|^2}{\sum_{m_1\in\mathcal{M}_1}{d}_{m_1}^{-{\alpha}}|{h}_{n,m_1}^{\text{Rician}}|^2},\ \ \ n\in \mathcal{N}\backslash\{1\}.
\label{iv23}
\end{aligned}
\end{equation}

Similar to the analysis for the swarm head, we first consider the numerator of \eqref{iv23}, i.e., $\overline{Y} \triangleq \big|\sum_{m_0\in\mathcal{M}_0}{d}_{m_0}^{\ \frac{-{\alpha}}{2}}{h}_{n,m_0}^{\text{Rician}}h^*_{1,{m_0}}/|h_{1,{m_0}}|\big|^2$. Note that available GBSs' beamforming weights $\omega_{m_0}=h^*_{1,{m_0}}/|h_{1,{m_0}}|$, $m_0\in\mathcal{M}_0$ are designed based on the channels with the swarm head, thus they are independent of the channels with the swarm members and will not affect the distribution of $\overline{Y}$. But in general, the distribution of the summation of random phase elements, i.e., $\overline{Y}$, is hard to find, because the magnitude of each element inside $\overline{Y}$ is random \cite{abdi2000pdf}. Therefore, we invoke the Central Limit Theorem, and the summation inside $\overline{Y}$ is approximated by a CSCG distributed random variable with zero mean and variance $\overline{Y}_{int} \triangleq \sum_{m_0\in\mathcal{M}_0}{d}_{m_0}^{-{\alpha}}$.

The remaining problem is to obtain the distribution of $\overline{Y}_{int}$. Instead of the previously used Gamma distribution, we adopt the Pearson type V distribution, i.e., the inverse Gamma distribution \cite{martin2017modeling}, to approximate the distribution of $\overline{Y}_{int}$, due to the following two reasons. First, it can provide a tight approximation as the Gamma distribution does. Second, it facilitates our subsequent analysis, as elaborated next.

\underline{\emph{Lemma 5.3}}: \ The PDF of $\overline{Y}_{int} = \sum_{m_0\in\mathcal{M}_0}{d}_{m_0}^{-{\alpha}}$ is approximated by
\begin{equation}
\begin{aligned}
f_{\overline{Y}_{int}}(\overline{y}_{int}) = \frac{b_{\overline{Y}}^{(a_{\overline{Y}})}}{\Gamma(a_{\overline{Y}})}\overline{y}_{int}^{(-a_{\overline{Y}}-1)}e^{-\frac{b_{\overline{Y}}}{\overline{y}_{int}}}, \ \ \ \overline{y}_{int}\geq 0, \nonumber
\end{aligned}
\end{equation}where $a_{\overline{Y}}$ and $b_{\overline{Y}}$ are the parameters of the inverse Gamma distribution. Therefore, the \eqref{iv23}'s numerator is exponentially distributed with mean $\overline{Y}_{int}$, i.e.,
\begin{equation}
\begin{aligned}
F_{\overline{Y}}(\overline{y})& = \int_{0}^{\infty}\Big(1-e^{-\frac{\overline{y}}{\overline{y}_{int}}}\Big)f_{\overline{Y}_{int}}(\overline{y}_{int})\,d\overline{y}_{int}\\
& = 1-\frac{b_{\overline{Y}}^{(a_{\overline{Y}})}}{\Gamma(a_{\overline{Y}})}\int_{0}^{\infty}\overline{y}_{int}^{(-a_{\overline{Y}}-1)}e^{-\big(b_{\overline{Y}}+\overline{y}\big)\frac{1}{\overline{y}_{int}}}\,d\overline{y}_{int}\\
& = 1-\bigg(\frac{b_{\overline{Y}}}{b_{\overline{Y}}+\overline{y}}\bigg)^{a_{\overline{Y}}}, \ \ \ \overline{y}\geq 0.
\label{iv24}
\end{aligned}
\end{equation}
\begin{IEEEproof}
See Appendix D.
\end{IEEEproof}
Now we elaborate the reason why the inverse Gamma distribution is used in Lemma 5.3. This is because it can introduce favourable $e^{-\frac{1}{\overline{y}_{int}}}$ term and make the final CDF in \eqref{iv24} to be a simple polynomial, which greatly eases our further analysis.

Next, we focus on the denominator of \eqref{iv23}, i.e., $\underline{Y} \triangleq \sum_{m_1\in\mathcal{M}_1}{d}_{m_1}^{-{\alpha}}|{h}_{n,m_1}^{\text{Rician}}|^2$, and clearly it has the same approximated distribution as \eqref{iv21}, i.e.,
\begin{equation}
\begin{aligned}
F_{\underline{Y}}(\underline{y}) = \frac{\gamma\big(a_{\underline{Y}},b_{\underline{Y}}\underline{y}\big)}{\Gamma(a_{\underline{Y}})}, \ \ \ \underline{y}\geq 0,
\label{iv25}
\end{aligned}
\end{equation}
with the parameters $a_{\underline{Y}}=a_{\underline{X}}$ and $b_{\underline{Y}}=b_{\underline{X}}$.

Given Lemma 5.3 and the distribution of \eqref{iv23}'s denominator $\underline{Y}$ in \eqref{iv25}, we have the following proposition.

\underline{\emph{Proposition 5.2}}: \ The probability that the swarm members can decode the common control message in Phase I is
\begin{equation}
\begin{aligned}
\mathcal{P}^{(\text{I-M})}(\theta^{(\text{I})})& = \mathbb{P}\bigg\{{\tt{SINR}_{n}^{(\text{I})}}=\frac{\overline{Y}}{\underline{Y}}\geq\theta^{(\text{I})}\bigg\} = 1 - F_{\frac{\overline{Y}}{\underline{Y}}}(\theta^{(\text{I})})\\
& = 1-\int_{0}^{\infty}\bigg(\int_{0}^{\theta^{(\text{I})}\underline{y}}f_{\overline{Y}}(\overline{y})\,d\overline{y}\bigg)f_{\underline{Y}}(\underline{y})\,d\underline{y}\\
& = 1-\bigg(1-\frac{b_{\underline{Y}}^{(a_{\underline{Y}})}}{\Gamma(a_{\underline{Y}})}\int_{0}^{\infty}\bigg(\frac{b_{\overline{Y}}}{b_{\overline{Y}}+\theta^{(\text{I})}\underline{y}}\bigg)^{a_{\overline{Y}}}\underline{y}^{(a_{\underline{Y}}-1)}e^{-b_{\underline{Y}}\underline{y}}\,d\underline{y}\bigg)\\
& \overset{\text{(a)}}{=} \bigg(\frac{b_{\underline{Y}}b_{\overline{Y}}}{\theta^{(\text{I})}}\bigg)^{a_{\underline{Y}}}\times\frac{1}{\Gamma(a_{\underline{Y}})}\int_{0}^{\infty}(1+t)^{-a_{\overline{Y}}}t^{a_{\underline{Y}}-1}e^{-\frac{b_{\underline{Y}}b_{\overline{Y}}}{\theta^{(\text{I})}}t}\,dt\\
& = \bigg(\frac{b_{\underline{Y}}b_{\overline{Y}}}{\theta^{(\text{I})}}\bigg)^{a_{\underline{Y}}}\Psi\bigg(a_{\underline{Y}},1+a_{\underline{Y}}-a_{\overline{Y}};\frac{b_{\underline{Y}}b_{\overline{Y}}}{\theta^{(\text{I})}}\bigg), \ \ n\in \mathcal{N}\backslash\{1\}.
\label{iv26}
\end{aligned}
\end{equation}
where $\overset{\text{(a)}}{=}$ follows by denoting $t=\frac{\theta^{(\text{I})}}{b_{\overline{Y}}}\underline{y}$, and $\Psi(a,b;z)=\frac{1}{\Gamma(a)}\int_{0}^{\infty}(1+t)^{b-a-1}t^{a-1}e^{-zt}\,dt$ is the Tricomi confluent geometric function [30, 6.5.(2)].

Given the probabilities that the swarm head and swarm members can decode the message in Phase I in Propositions 5.1 and 5.2, and thanks to the linearity of expectation, we have the following theorem.

\underline{\emph{Theorem 5.1}}: \ The expected number of UAVs that can decode the common control message in Phase I is
\begin{equation}
\begin{aligned}
\mathbb{E}_{\{\bold{d}, \bold{H}^{\text{Rician}}\}}\big\{\big|\mathbf{\Theta}^{(\text{I})}\big|\big\}& =
\sum_{n\in\mathcal{N}}\mathbb{P}\big\{{\tt{SINR}_{n}^{(\text{I})}}\geq\theta^{(\text{I})}\big\}\\
& = \mathcal{P}^{(\text{I-H})}(\theta^{(\text{I})}) + (N-1)\mathcal{P}^{(\text{I-M})}(\theta^{(\text{I})}).
\label{iv27}
\end{aligned}
\end{equation}

\section{Analysis of Phase II}

In this section, we aim to characterize the expected number of UAVs that can decode the common control message in Phase II, given that $|\mathbf{\Theta}^{(\text{I})}|=K$ UAVs have decoded it in Phase I, i.e., $\mathbb{E}_{\{\tilde{\bold{D}}, \bold{G}^{\text{Rayleigh}}\}}\{|\mathbf{\Theta}^{(\text{II})}|\big||\mathbf{\Theta}^{(\text{I})}|=K\}$. Similarly to the previous section, we first obtain the probability that a receiving UAV can decode the message in Phase II given that $K$ UAVs have decoded it in Phase I, and then apply the linearity of expectation to calculate the expected number of UAVs that can decode the message in Phase II.

The SINR at receiving UAV $n$ in Phase II given in \eqref{iii13} can be written as
\begin{equation}
\begin{aligned}
{\tt{SINR}_{n}^{(\text{II})}} = \frac{\tilde{P}\big|\sum_{k\in\mathbf{\Theta}^{(\text{I})}}g_{n,k}\big|^2}{\tilde{\sigma}^{2}}
= \frac{\tilde{P}\tilde{\beta}\Big|\sum_{k\in\mathbf{\Theta}^{(\text{I})}}\tilde{d}_{n,k}^{\ \frac{-\tilde{\alpha}}{2}}g_{n,k}^{\text{Rayleigh}}\Big|^2}{\tilde{\sigma}^{2}},\ \ \ n\in \mathcal{N}\backslash\mathbf{\Theta}^{(\text{I})}.
\label{iv28}
\end{aligned}
\end{equation}It directly follows that $Z \triangleq \big|\sum_{k\in\mathbf{\Theta}^{(\text{I})}}\tilde{d}_{n,k}^{\ \frac{-\tilde{\alpha}}{2}}g_{n,k}^{\text{Rayleigh}}\big|^2$ is exponentially distributed with mean $Z_{int} \triangleq \sum_{k\in\mathbf{\Theta}^{(\text{I})}}\tilde{d}_{n,k}^{-\tilde{\alpha}}$. The next step is to use the Pearson type V distribution for approximating the distribution of $Z_{int}$.

\begin{figure}[h]
  \centering
  \subfigure[Actual model.]{
    \includegraphics[width=2.0in]{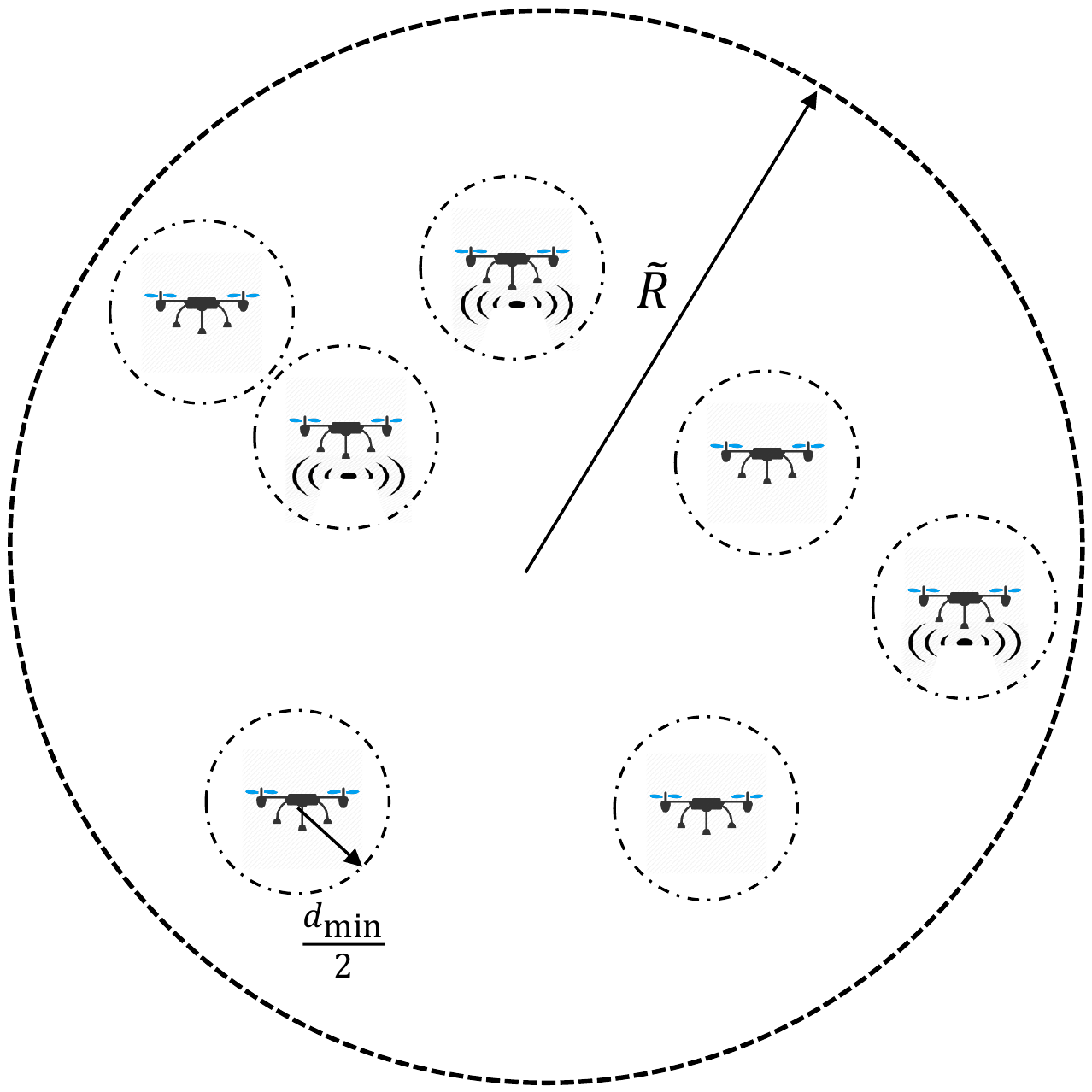}}
  \hspace{0.8in}
  \subfigure[Simplified model.]{
    \includegraphics[width=2.0in]{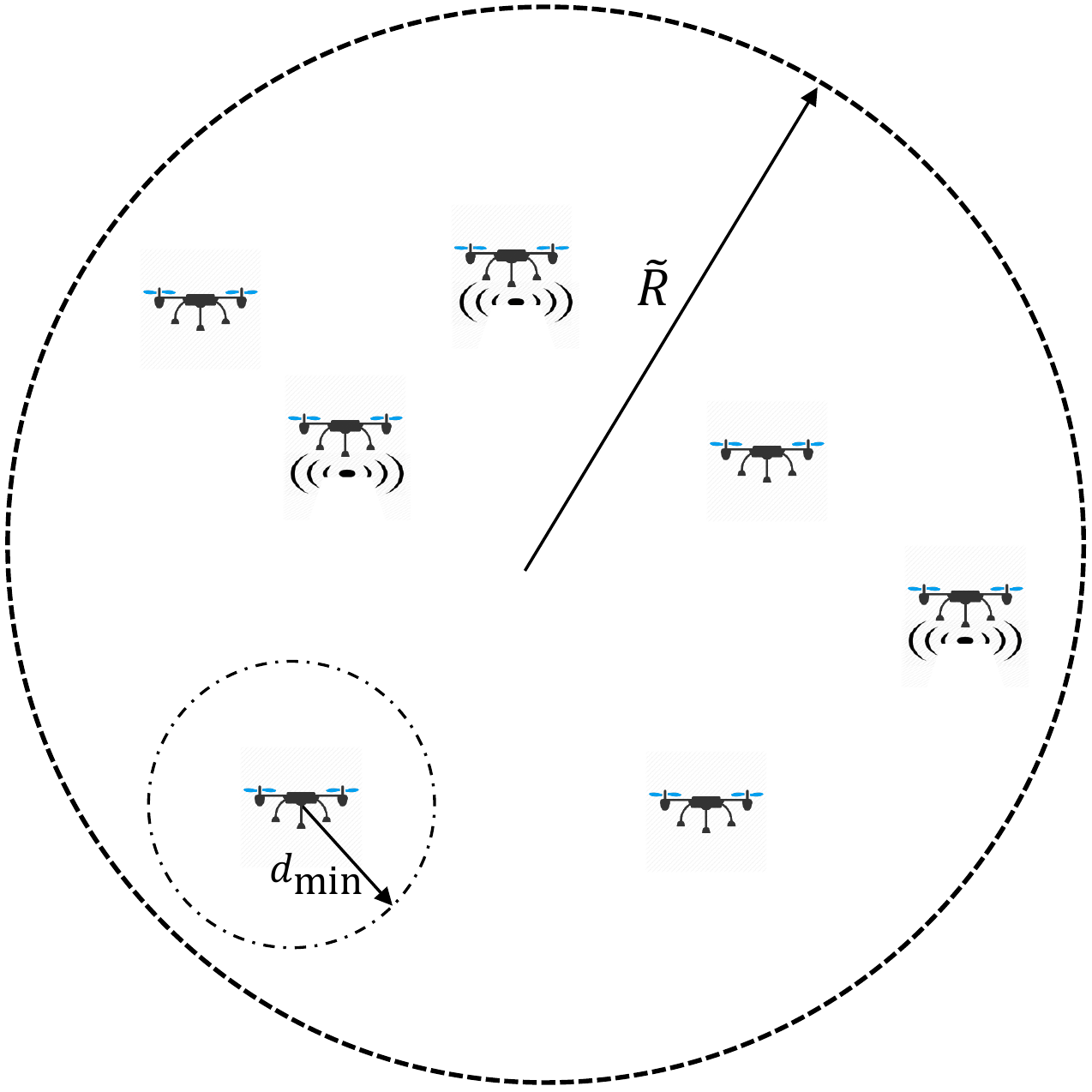}}
  \caption{Two UAV models in Phase II.}
  \label{smp2}
\end{figure}

Recall that $N$ UAVs are randomly located inside a circular horizontal disk with radius $\tilde{R}$ and altitude $H$, following a ``hard-core'' point process with a minimal separation distance $d_{\min}$ between any two UAVs \cite{haenggi2011mean}. Thus, the distances between UAVs, $\tilde{d}_{n,k}$'s, are dependent, because one UAV's location constrains those of the rest $N-1$ UAVs in the swarm. Equivalently, each UAV can be represented by a circle with radius $d_{\min}/2$ and all the circles do not overlap with each other, as shown in Fig.~\ref{smp2} (a).

According to \cite{haenggi2011mean}, the joint distribution of $\tilde{d}_{n,k}$'s is impractical to obtain, but by assuming the radius of the UAV swarm $\tilde{R}$ is much larger than the minimal separation distance $d_{\min}$, the probability that the distance between two UAVs is smaller than $d_{\min}$ is small. Thus, we can adopt the simplified model with the locations of UAVs following a modified BPP, i.e., $N$ UAVs are randomly located inside the swarm with radius $\tilde{R}$; besides, while a certain receiving UAV $n$ is viewed as a circle with radius $d_{\min}$, the other UAVs are viewed as points, and these points do not fall into the circle, as shown in Fig.~\ref{smp2} (b). By doing so, we can restore the independency of transmitting UAVs' locations, i.e.,
\begin{equation}
\begin{aligned}
f_{W_{n,1},\cdots,W_{n,K}}(w_{n,1},\cdots,w_{n,K}) \approx \prod_{k\in\mathbf{\Theta}^{(\text{I})}}f_{W_{n,k}}(w_{n,k}),\ \ \ n\in \mathcal{N}\backslash\mathbf{\Theta}^{(\text{I})},
\label{iv29}
\end{aligned}
\end{equation}
where $f_{W_{n,k}}(w_{n,k})$ is the PDF of the distance $\tilde{d}_{n,k}$ between transmitting UAV $k$ and receiving UAV $n$.

\underline{\emph{Lemma 6.1}}: \ Given that $K$ UAVs have decoded the message in Phase I, the CDF of $Z = \big|\sum_{k\in\mathbf{\Theta}^{(\text{I})}}\tilde{d}_{n,k}^{\ \frac{-\tilde{\alpha}}{2}}g_{n,k}^{\text{Rayleigh}}\big|^2$ is approximated by
\begin{equation}
\begin{aligned}
F_{Z}(z)& = \int_{0}^{\infty}\Big(1-e^{-\frac{z}{z_{int}}}\Big)f_{Z_{int}}(z_{int})\,dz_{int}\\
& = 1-\bigg(\frac{b_{Z}(K)}{b_{Z}(K)+z}\bigg)^{a_{Z}(K)}, \ \ \ z\geq 0.
\label{iv30}
\end{aligned}
\end{equation}
\begin{IEEEproof}
See Appendix E.
\end{IEEEproof}

Given Lemma 6.1, the probability that a receiving UAV can decode the message in Phase II given that $K$ UAVs have decoded it in Phase I is
\begin{equation}
\begin{aligned}
\mathcal{P}^{(\text{II})}(\theta^{(\text{II})},K)& = \mathbb{P}\bigg\{{\tt{SINR}_{n}^{(\text{II})}}=\frac{\tilde{P}\tilde{\beta}Z}{\tilde{\sigma}^{2}}\geq\theta^{(\text{II})}\bigg\} = 1 - F_{Z}\bigg(\frac{\tilde{\sigma}^{2}\theta^{(\text{II})}}{\tilde{P}\tilde{\beta}}\bigg)\\
& = \bigg(\frac{b_{Z}(K)}{b_{Z}(K)+\frac{\tilde{\sigma}^{2}\theta^{(\text{II})}}{\tilde{P}\tilde{\beta}}}\bigg)^{a_{Z}(K)},\ \ \ n\in \mathcal{N}\backslash\mathbf{\Theta}^{(\text{I})}.
\label{iv31}
\end{aligned}
\end{equation}Thanks to the linearity of expectation, we can obtain the expected number of UAVs that can decode the common control message in Phase II given that $K$ UAVs have decoded it in Phase I, i.e.,
\begin{equation}
\begin{aligned}
\mathbb{E}_{\{\tilde{\bold{D}}, \bold{G}^{\text{Rayleigh}}\}}\Big\{\big|\mathbf{\Theta}^{(\text{II})}\big|\Big|\big|\mathbf{\Theta}^{(\text{I})}\big|=K\Big\}& =
\sum_{n\in \mathcal{N}\backslash\mathbf{\Theta}^{(\text{I})}}\mathbb{P}\big\{{\tt{SINR}_{n}^{(\text{II})}}\geq\theta^{(\text{II})}\big\}\\
& = \big(N-K\big)\mathcal{P}^{(\text{II})}\big(\theta^{(\text{II})},K\big).
\label{iv32}
\end{aligned}
\end{equation}

With Theorem 5.1 and \eqref{iv32}, \eqref{iv18} can be rewritten as
\begin{equation}
\begin{aligned}
\eta&\approx\frac{1}{N}\Big(\mathbb{E}_{\{\bold{d}, \bold{H}^{\text{Rician}}\}}\big\{\big|\mathbf{\Theta}^{(\text{I})}\big|\big\}+\sum_{K=1}^{N}\mathbb{P}\big\{\big|\mathbf{\Theta}^{(\text{I})}\big|=K\big\}\times\mathbb{E}_{\{\tilde{\bold{D}}, \bold{G}^{\text{Rayleigh}}\}}\Big\{\big|\mathbf{\Theta}^{(\text{II})}\big|\Big|\big|\mathbf{\Theta}^{(\text{I})}\big|=K\Big\}\Big),
\label{add1}
\end{aligned}
\end{equation}
and the final task is to calculate the probability that $K$ UAVs can decode the common control message in Phase I, i.e., $\mathbb{P}\{|\mathbf{\Theta}^{(\text{I})}|=K\}$.

\begin{figure}[!t]
\centering
\includegraphics[width=3.0in]{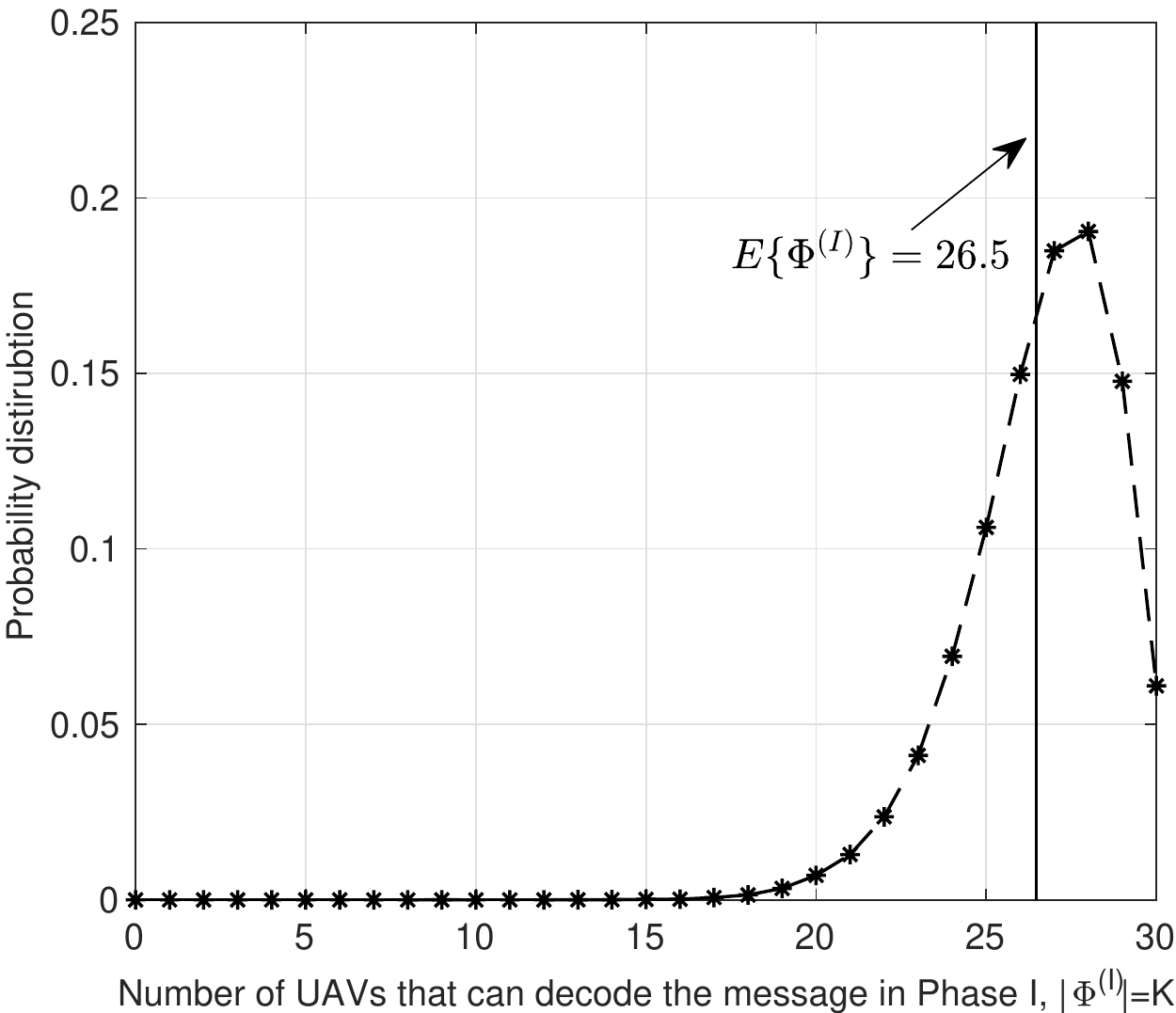}
\caption{The probability distribution of the number of UAVs that can decode the message in Phase I, $|\mathbf{\Theta}^{(\text{I})}|$, in the low SINR requirement regime ($\theta^{\text{(I)}}=0.25$), with $M_0=8$, $M_1=4$, $N=30$, $R=900$ m, $\tilde{R}=30$ m, $H=300$ m, and $d_{\min}=5$ m. Other parameters are given in Table I.}
\label{k}
\end{figure}

It is worth noting that from a GBS's point of view, its channels to all the UAVs in the swarm have nearly the same path loss, as in Assumption 4.1, therefore the channels from one GBS to all the UAVs in the swarm are uncorrelated but not independent. As a result, the SINRs at UAVs in Phase I given in \eqref{iii6} and \eqref{iii7} are also not independent, and the probability that $K$ UAVs can decode the message in Phase I, $\mathbb{P}\{|\mathbf{\Theta}^{(\text{I})}|=K\}$, cannot be simply represented by the product of the probabilities that the swarm head and swarm members can decode the message in Phase I as in \eqref{iv22} and \eqref{iv26}.

Nevertheless, the size of common control message is generally small, which allows us to focus on the low SINR requirement regime (e.g., $\theta^{(\text{I})}=0.25$ as shown in Fig.~\ref{k}), where the reliability is relatively high. In this regime, a large portion of UAVs (around $90\%$ in Fig.~\ref{k}) can already decode the message in Phase I. Thus, the distribution of $|\mathbf{\Theta}^{(\text{I})}|$ centers on $\mathbb{E}_{\{\bold{d}, \bold{H}^{\text{Rician}}\}}\{|\mathbf{\Theta}^{(\text{I})}|\}$, and we can use $\mathbb{E}_{\{\tilde{\bold{D}}, \bold{G}^{\text{Rayleigh}}\}}\big\{|\mathbf{\Theta}^{(\text{II})}|\big|\mathbb{E}_{\{\bold{d}, \bold{H}^{\text{Rician}}\}}\{|\mathbf{\Theta}^{(\text{I})}|\}\big\}$ to approximate $\sum_{K=1}^{N}\mathbb{P}\{|\mathbf{\Theta}^{(\text{I})}|=K\}\times\mathbb{E}_{\{\tilde{\bold{D}}, \bold{G}^{\text{Rayleigh}}\}}\big\{|\mathbf{\Theta}^{(\text{II})}|\big||\mathbf{\Theta}^{(\text{I})}|=K\big\}$. Therefore, \eqref{add1} can be further simplified and we have the following theorem.

\underline{\emph{Theorem 6.1}}: \  In our interested low SINR requirement regime, the expected percentage of UAVs in the swarm that can decode the common control message with the proposed two-phase transmission protocol is approximated as
\begin{equation}
\begin{aligned}
\eta
&\approx\frac{1}{N}\Big(\mathbb{E}_{\{\bold{d}, \bold{H}^{\text{Rician}}\}}\big\{\big|\mathbf{\Theta}^{(\text{I})}\big|\big\}+\mathbb{E}_{\{\tilde{\bold{D}}, \bold{G}^{\text{Rayleigh}}\}}\Big\{\big|\mathbf{\Theta}^{(\text{II})}\big|\Big|\mathbb{E}_{\{\bold{d}, \bold{H}^{\text{Rician}}\}}\big\{\big|\mathbf{\Theta}^{(\text{I})}\big|\big\}\Big\}\Big),
\label{iv33}
\end{aligned}
\end{equation}
where $\mathbb{E}_{\{\tilde{\bold{D}}, \bold{G}^{\text{Rayleigh}}\}}\big\{|\mathbf{\Theta}^{(\text{II})}|\big|\mathbb{E}_{\{\bold{d}, \bold{H}^{\text{Rician}}\}}\{|\mathbf{\Theta}^{(\text{I})}|\}\big\}$ is given in \eqref{iv32} with $K=\mathbb{E}_{\{\bold{d}, \bold{H}^{\text{Rician}}\}}\{|\mathbf{\Theta}^{(\text{I})}|\}$, and $\mathbb{E}_{\{\bold{d}, \bold{H}^{\text{Rician}}\}}\{|\mathbf{\Theta}^{(\text{I})}|\}$ is given in \eqref{iv27}.

\section{Numerical Results}

\begin{table*}[t]
\newcommand{\tabincell}[2]{\begin{tabular}{@{}#1@{}}#2\end{tabular}}
\centering
\caption{Simulation Parameters}
\begin{tabular}{|c|c|}
\hline
Path loss exponent in Phase I and Phase II: ${\alpha}$ and $\tilde{\alpha}$ & $2$\\
\hline
Rician factor: $\kappa$ & $4$\\
\hline
SINR gap in Phase I and Phase II: $\rho$ and $\tilde{\rho}$ & $5/6$\\
\hline
Transmission power of GBSs: $P$ & $43$ dBm\\
\hline
Transmission power of UAVs: $\tilde{P}$ & $23$ dBm\\
\hline
Transmission bandwidth in Phase I and Phase II: $B$ and $\tilde{B}$ & $200$ KHz\\
\hline
Interference plus noise power level at receiving UAVs in Phase II: $\tilde{\sigma}^{2}$ & $-40$ dBm\\
\hline
Transmission time of Phase I and Phase II: $\tau^{(\text{I})}$ and $\tau^{(\text{II})}$ & $0.5$ ms\\
\hline
\end{tabular}
\end{table*}

We conduct simulations with GBSs randomly located inside a circular ground area with radius ${R}=900$ m, and UAVs are randomly located inside a circular aerial area with radius $\tilde{R}=30$ m and height $H=300$ m. The minimal separation distance between any two UAVs is $d_{\min}=5$ m. We set the carrier frequency of downlink cellular communication as $2$ GHz and that of D2D communication as $2.4$ GHz, therefore Phase I and Phase II have very similar channel power gain $\beta/\tilde{\beta}$ at the reference distance ${d}_{\text{ref}}/\tilde{d}_{\text{ref}}= 1$ m. For simplicity, we set them to be the same, i.e., $\beta=\tilde{\beta}=-40$ dB. Other parameters are given in Table I. Here, $100,000$ random locations of GBSs and UAVs as well as channel realizations in Phase I and Phase II are generated to calculate the reliability performance in terms of $\eta$. Note that in order to demonstrate and compare the reliability performance $\eta$ more clearly, we set y-axis to be log-scaled and plot $1-\eta$ instead of $\eta$ when it is needed. This implies lower the $1-\eta$, higher the reliability performance.

\subsection{Verification of Analytical Results}

\begin{figure*}[t]
\centering
\subfigure[Reliability performance after Phase I, $1-\frac{\mathbb{E}\{|\mathbf{\Theta}^{(\text{I})}|\}}{N}$, versus the size of common control message, $D$, under different numbers of available and occupied GBSs $(M_0,M_1)$'s.]{
\includegraphics[width=2.0in]{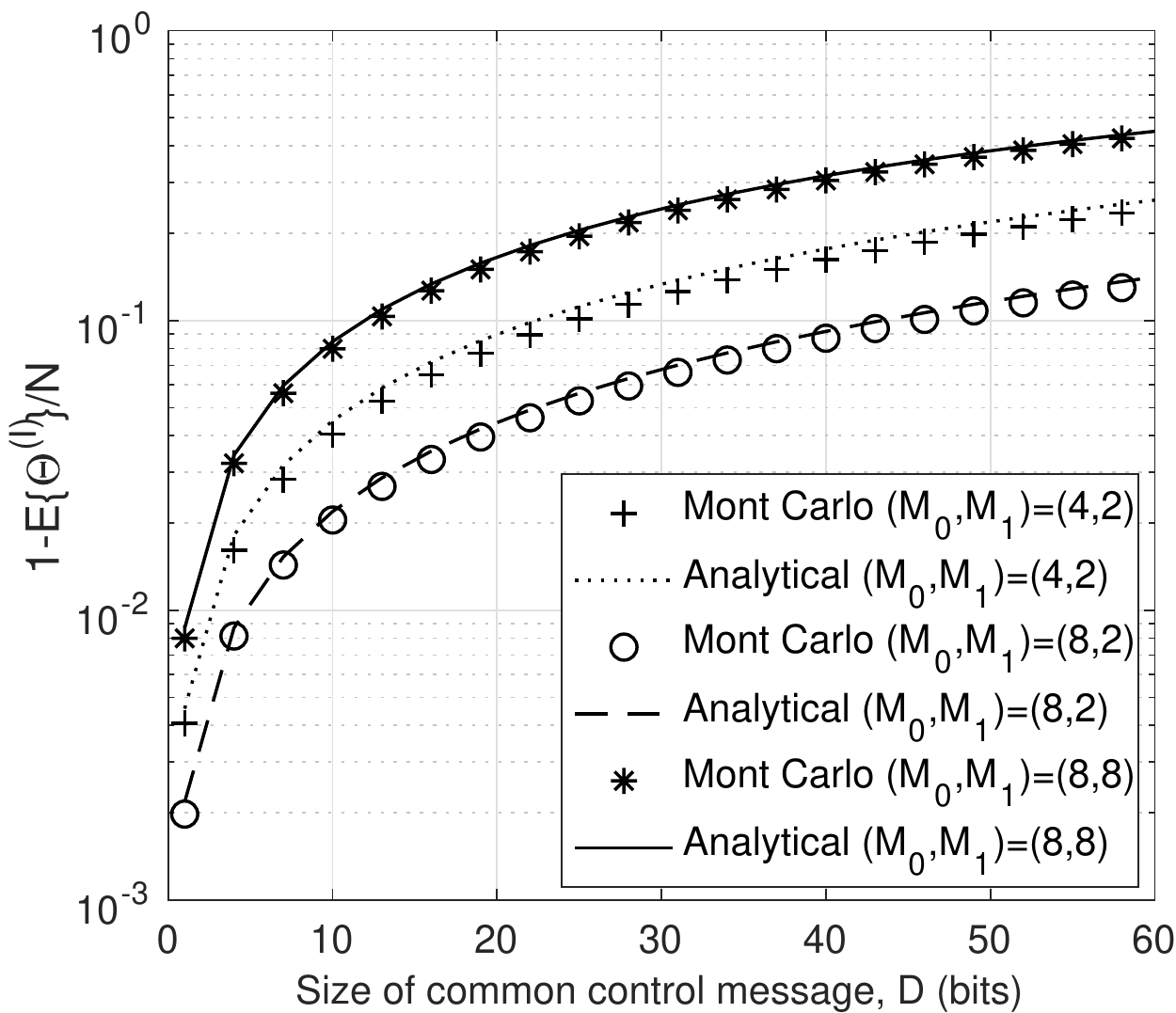}}
\hspace{0.05in}
\subfigure[Reliability performance (after Phase II), $1-\eta$, versus the size of common control message, $D$, under different numbers of available and occupied GBSs $(M_0,M_1)$'s.]{
\includegraphics[width=2.0in]{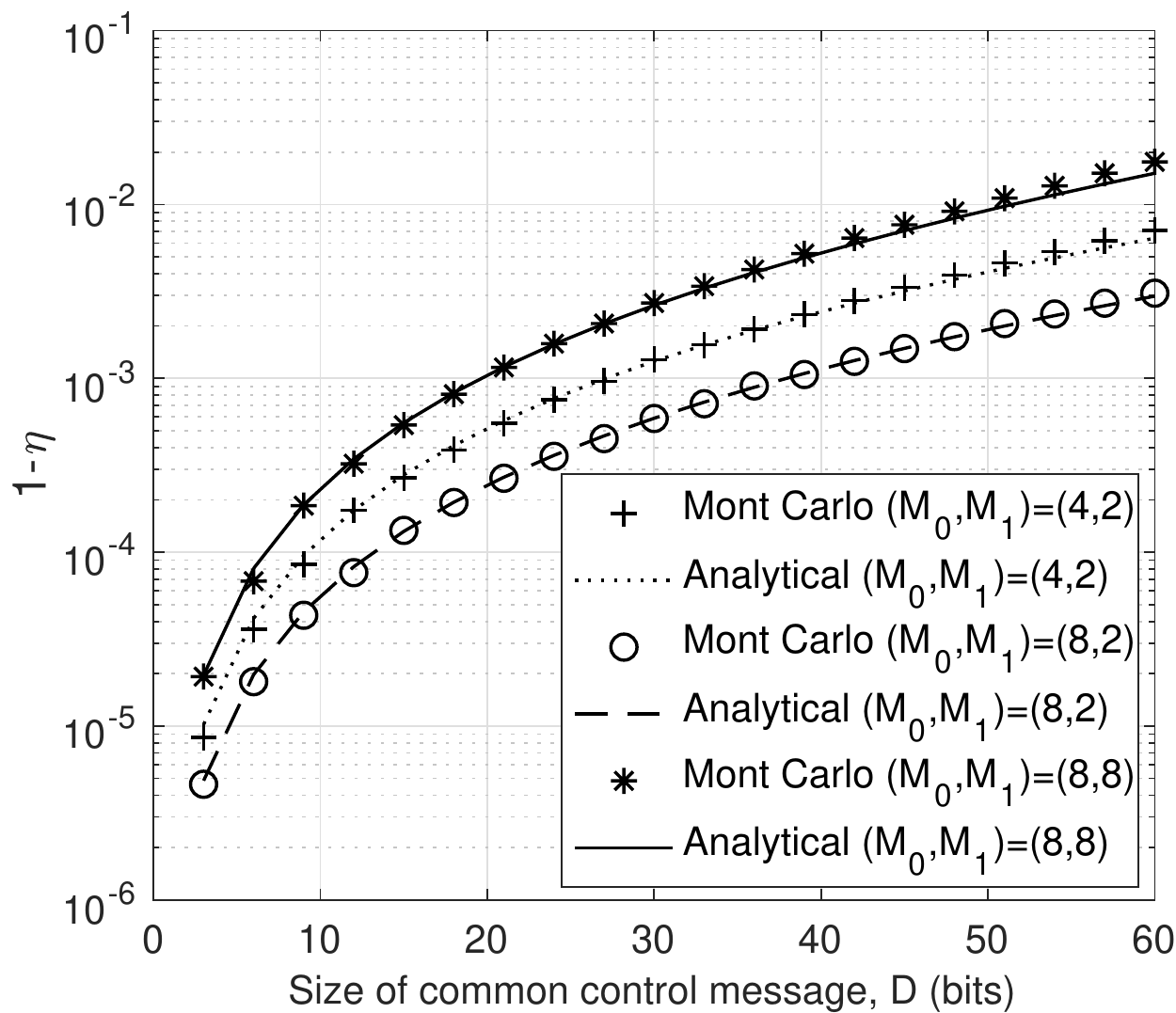}}
\hspace{0.05in}
\subfigure[Reliability performance, $1-\eta$, versus the size of common control message, $D$, under different numbers of UAVs $N$'s.]{
\includegraphics[width=2.0in]{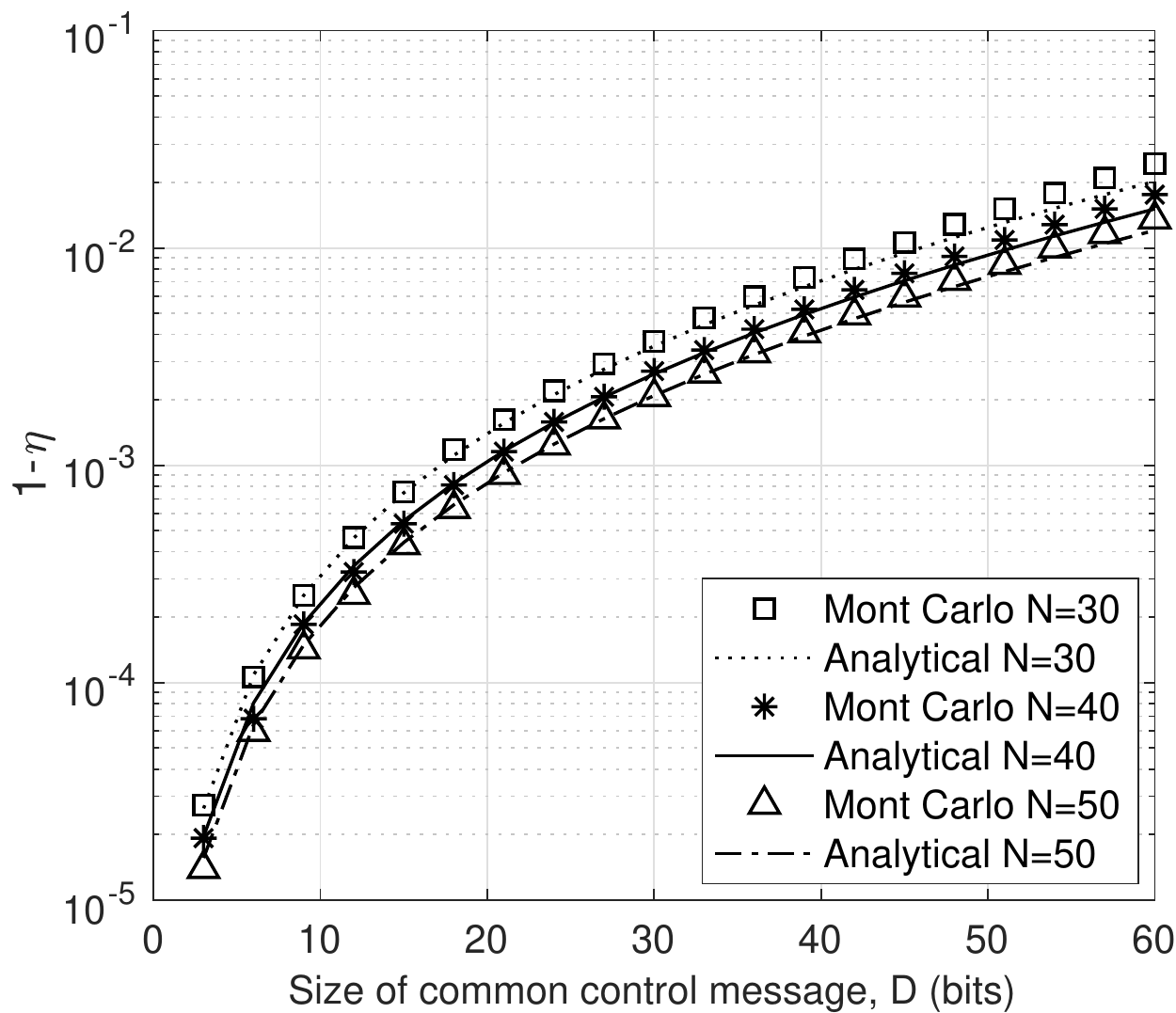}}
\caption{Monte Carlo simulations to verify analytical results.}
\label{mva2}
\vspace{-1em}
\end{figure*}

In this subsection, we verify the accuracy of our analytical results under different setups. Note that we only focus on the low SINR requirement region, where the reliability performance is relatively high.

Fig.~\ref{mva2} (a) plots the expected percentage of UAVs that cannot decode the message after Phase I, i.e., $1-\frac{\mathbb{E}\{|\mathbf{\Theta}^{(\text{I})}|\}}{N}$, while Fig.~\ref{mva2}(b) plots the expected percentage of UAVs that cannot decode the message after Phase II, i.e., $1-\eta$, over the size of common control message, $D$ (bits), under different numbers of available and occupied GBSs $(M_0,M_1)$'s, with totally $N=40$ UAVs in the swarm. Fig.~\ref{mva2} (c) plots $1-\eta$ over $D$, under different numbers of UAVs $N$'s, with $(M_0,M_1)=(8,8)$. Note that larger the size of message $D$, higher the SINR requirements $\theta^{\text{(I)}}$ and $\theta^{\text{(II)}}$ in two transmission phases, as in \eqref{iii8} and \eqref{iii14}. As we can see, despite all the assumptions and approximations, our analytical results match well with the Monte Carlo curves under different setups, indicating their accuracy in our interested low SINR requirement regime.

As shown in Fig.~\ref{mva2} (a), in the low SINR requirement regime, a large portion of UAVs (around 90\%) can already decode the message from the cellular network in Phase I. But since the ground-to-air channels are assumed to stay constant over the entire $\tau$ s, in order to achieve a high reliability performance, e.g., $\eta\geq99\%$ in Fig.~\ref{mva2} (b), Phase II is needed to create independent fading and provide diversity gain. We can also observe from Fig.~\ref{mva2} (a) that the reliability performance after Phase I depends on the ratio of the numbers of available and occupied GBSs, i.e., $\frac{M_0}{M_1}$, rather than their exact numbers $M_0$ and $M_1$. For example, the reliability performance after Phase I under $(M_0, M_1) = (8,2)$ is better than that under $(M_0, M_1) = (8,8)$, because the former gives a higher $\frac{M_0}{M_1}$, in turn higher average SINRs at UAVs in Phase I, and thus a more reliable reception in Phase I.

From Fig.~\ref{mva2} (c), we can see that the reliability performance increases as the number of UAVs $N$ increases. This is because although the expected percentage of UAVs that can decode the message in Phase I, $\frac{\mathbb{E}\{|\mathbf{\Theta}^{(\text{I})}|\}}{N}$, does not change with more number of UAVs $N$ in the swarm, the expected number of UAVs that can decode the message in Phase I, $\mathbb{E}\{|\mathbf{\Theta}^{(\text{I})}|\}$, increases, as in \eqref{iv27}, which means more UAVs can help the D2D communication in Phase II, and thus a more reliable reception in Phase II.

\subsection{Performance Comparison}

In this subsection, we aim to show the effectiveness of the studied two-phase transmission protocol. Three benchmark protocols are introduced for performance comparison.
\begin{itemize}
\item{Protocol I: Only the available GBS with the shortest distance to the UAV swarm transmits the common control message for the entire $\tau$ s, no D2D communication.}
\item{Protocol II: All the available GBSs transmit the common control message for the entire $\tau$ s, no D2D communication.}
\item{Protocol III: The same as our proposed protocol, except that only the swarm head helps relay the common control message in Phase II.}
\end{itemize}

\begin{figure*}[t]
\centering
\subfigure[$D=4$ bits.]{
\includegraphics[width=2.8in]{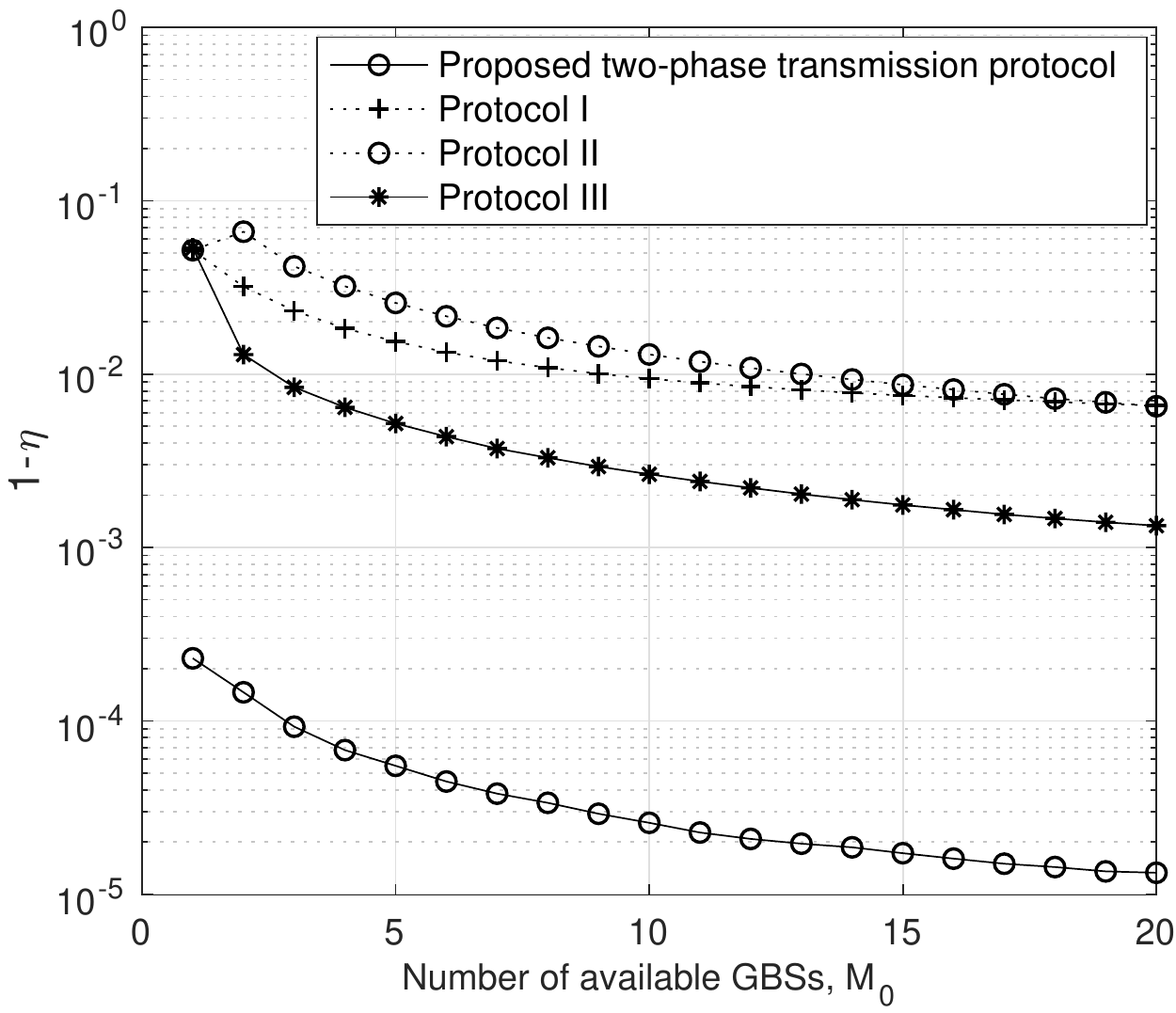}}
\hspace{0.1in}
\subfigure[$D=40$ bits.]{
\includegraphics[width=2.8in]{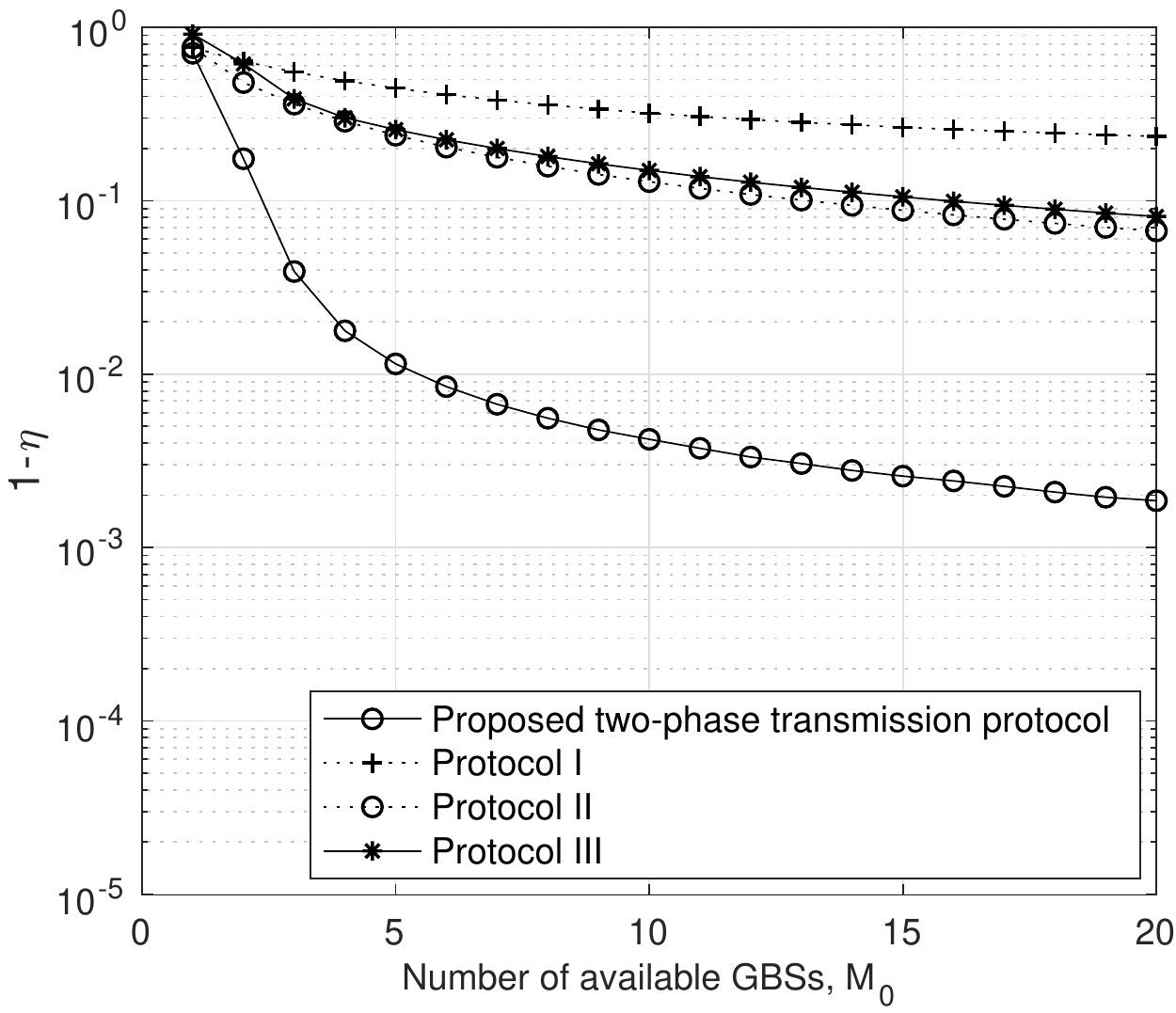}}
\caption{Performance comparison with benchmark protocols.}
\label{2v1}
\vspace{-1em}
\end{figure*}

Fig.~\ref{2v1} compares the reliability performance of our two-phase transmission protocol with three benchmark protocols, under different sizes of common control message $D$'s, with $M_1=8$ occupied GBSs and $N=40$ UAVs. It can be observed that our two-phase transmission protocol is the most reliable one. Besides that, we also have the following observations.

\begin{itemize}
\item{Comparing Protocol I with Protocol II, we can see that only with very small size of message, e.g., $D=4$ bits in Fig.~\ref{2v1} (a), letting the closest available GBS to transmit (Protocol I) is more reliable. Since the LoS component in Rician fading ground-to-air channels can guarantee a certain level of useful signal, while letting all the available GBSs to transmit (Protocol II) can cause signals non-coherently adding up at the UAVs and degenerate the reliability performance of cellular downlink communication. But as $D$ increases, the SINR requirement also increases, and the closest GBS quickly fails to support the UAV swarm, due to its limited power gain comparing with the strong interference from occupied GBSs, e.g., $D=40$ bits in Fig.~\ref{2v1} (b). While letting all the available GBSs to transmit at the same frequency band can provide a higher power gain to combat strong downlink interference, and thus a better reliability performance is achieved.}
\item{Comparing our proposed protocol with Protocol II, we can see that letting all the available GBSs to transmit (Protocol II) comes at the cost of some UAVs failing to decode the message due to the uncorrelated ground-to-air channels. Also the aforementioned channels are assumed to stay constant over the entire $\tau$ s, therefore Phase II is needed to create independent fading and provide extra diversity gain. The UAVs that failed to decode the message in cellular downlink communication can utilize this second chance to decode it with a high reliability, due to the less interfered D2D channels and the proximity among UAVs.}
\item{Comparing our proposed protocol with Protocol III, we can see that due to the Rayleigh fading D2D channels in Phase II, only letting the swarm head to help relay the message in Phase II (Protocol III) can cause some receiving UAVs to be in deep fading and fail to decode the message. While letting all the UAVs to help relay the message (our protocol) can provide a higher power gain in D2D communication, and greatly reduce the probability that the ``added-up" D2D channel at certain receiving UAV is in deep fading, therefore a better reliability performance can be achieved.}
\end{itemize}

\subsection{Effect of Swarm Head}

\begin{figure}[!t]
\centering
\includegraphics[width=2.8in]{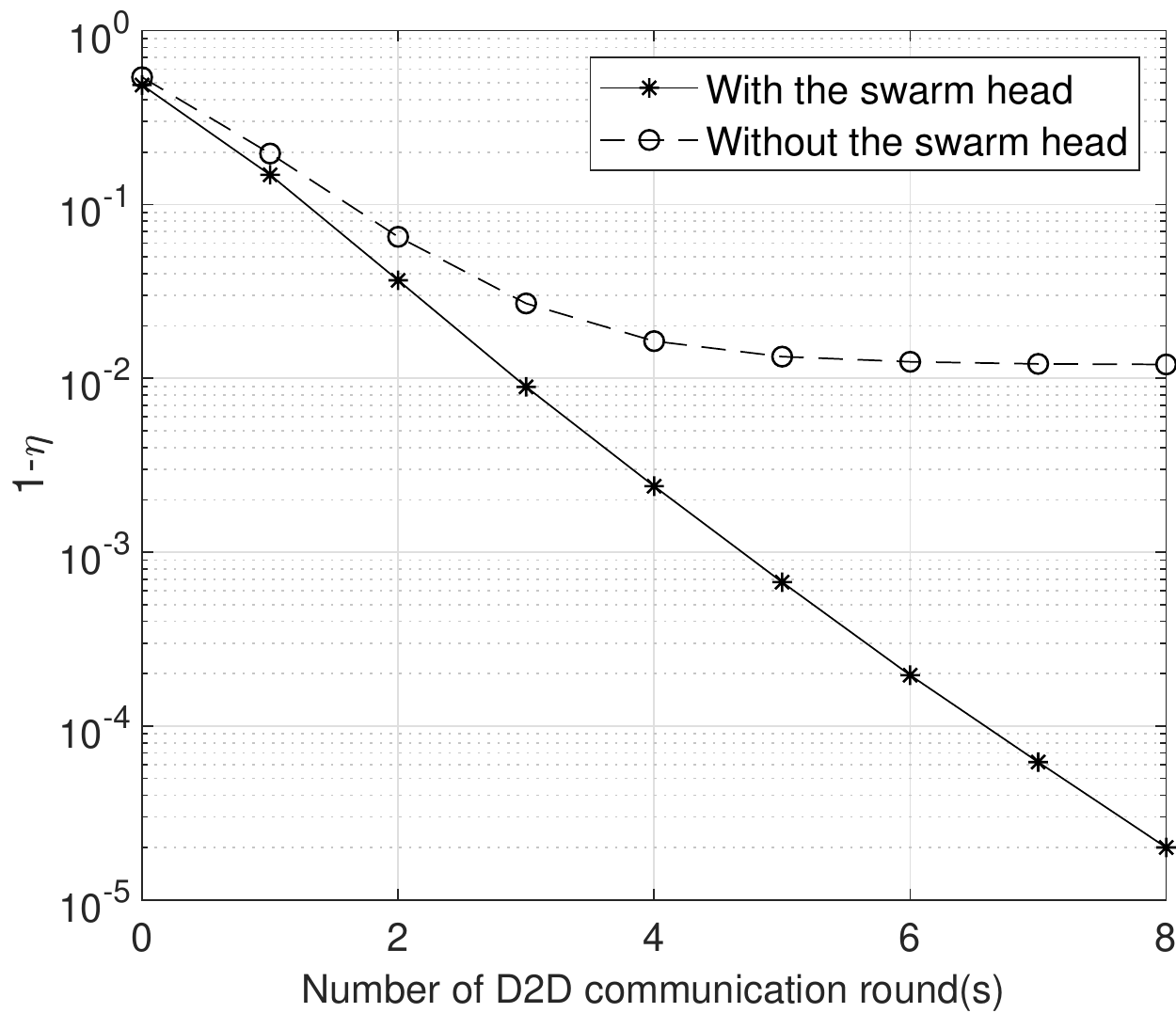}
\caption{Effect of the swarm head and number of D2D communication round(s).}
\label{head}
\end{figure}

The effect of the swarm head is subtle in our proposed two-phase protocol, when the number of UAVs $N$ is large. This is because a large proportion of UAVs, mostly swarm members, can already decode the message in Phase I in the low SINR requirement regime, as in Fig.~\ref{mva2} (a), and the presence of the swarm head makes little difference.

In this subsection, we explore the effect of the swarm head beyond the low SINR requirement regime, i.e., $D=150$ bits, with a small number of UAVs $N=10$, $M_0=8$ available GBSs, and $M_1=8$ occupied GBSs. Under this setup, our proposed two-phase transmission protocol can no longer provide a high reliability performance, therefore we relax the latency requirement and extend to the case with multiple phases. Specifically, after the cellular downlink communication with duration $\tau$ s, there are multiple rounds of D2D communication. Each round is of $\tau$ s, and all the UAVs that have decoded the message previously help relay it to the other UAVs in the swarm. We assume that the nominal values of the locations of UAVs and in turn the path loss of D2D channels do not change for the entire transmission, but the small-scale fading of D2D channels is independent over different D2D communication rounds (due to small perturbations of each UAV's location around its nominal value with respect to the carrier wavelength in practice).

Fig.~\ref{head} plots the reliability performance with and without the swarm head versus the number of D2D communication round(s). As we can see, the reliability performance with the swarm head increases fast as the number of D2D communication rounds increases, while that without the swarm head quickly converges. This is because with at least one UAV that can decode the message in cellular downlink communication, after sufficient rounds of D2D communication, all the UAVs in the swarm can decode the message. For the protocol with the swarm head, the swarm head can decode the message from cellular network with a very high probability due to dedicatedly designed beamforming, therefore a high reliability performance can be achieved with sufficient rounds of D2D communication. While for the protocol without the swarm head, all the UAVs opportunistically decode the message from cellular network, and the reliability performance only converges to around $99\%$.

\subsection{Effect of Key System Parameters}

In this subsection, we study the effect of key system parameters on the reliability performance of our proposed two-phase transmission protocol, with $M_0=8$ available GBSs, $M_1=8$ occupied GBSs, $N=40$ UAVs and $D=40$ bits common control message.

\begin{figure*}[t]
\centering
\subfigure[Effect of the radius of the UAV swarm $\tilde{R}$.]{
\includegraphics[width=2.0in]{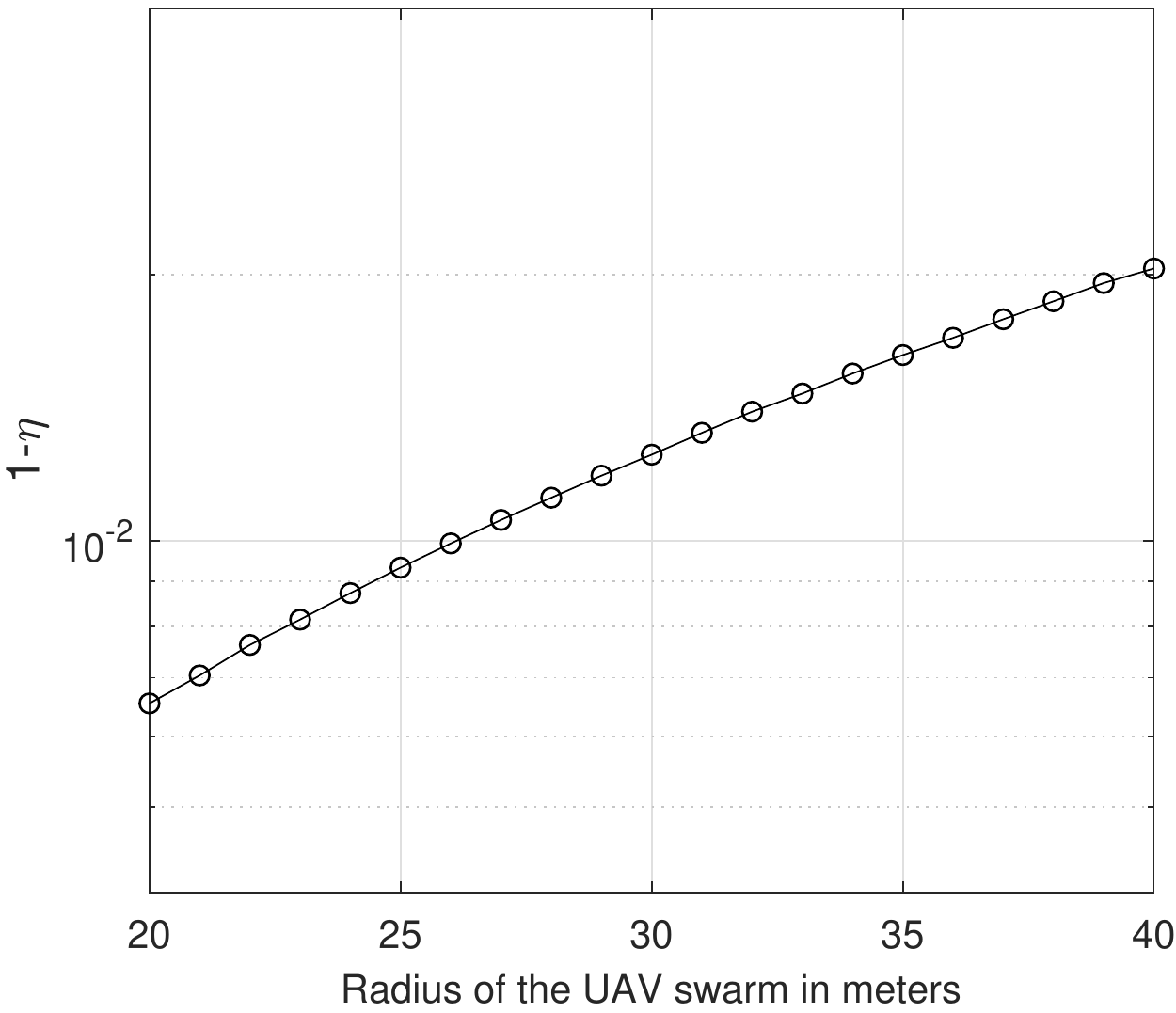}}
\hspace{0.05in}
\subfigure[Effect of the height of the UAV swarm $H$.]{
\includegraphics[width=2.0in]{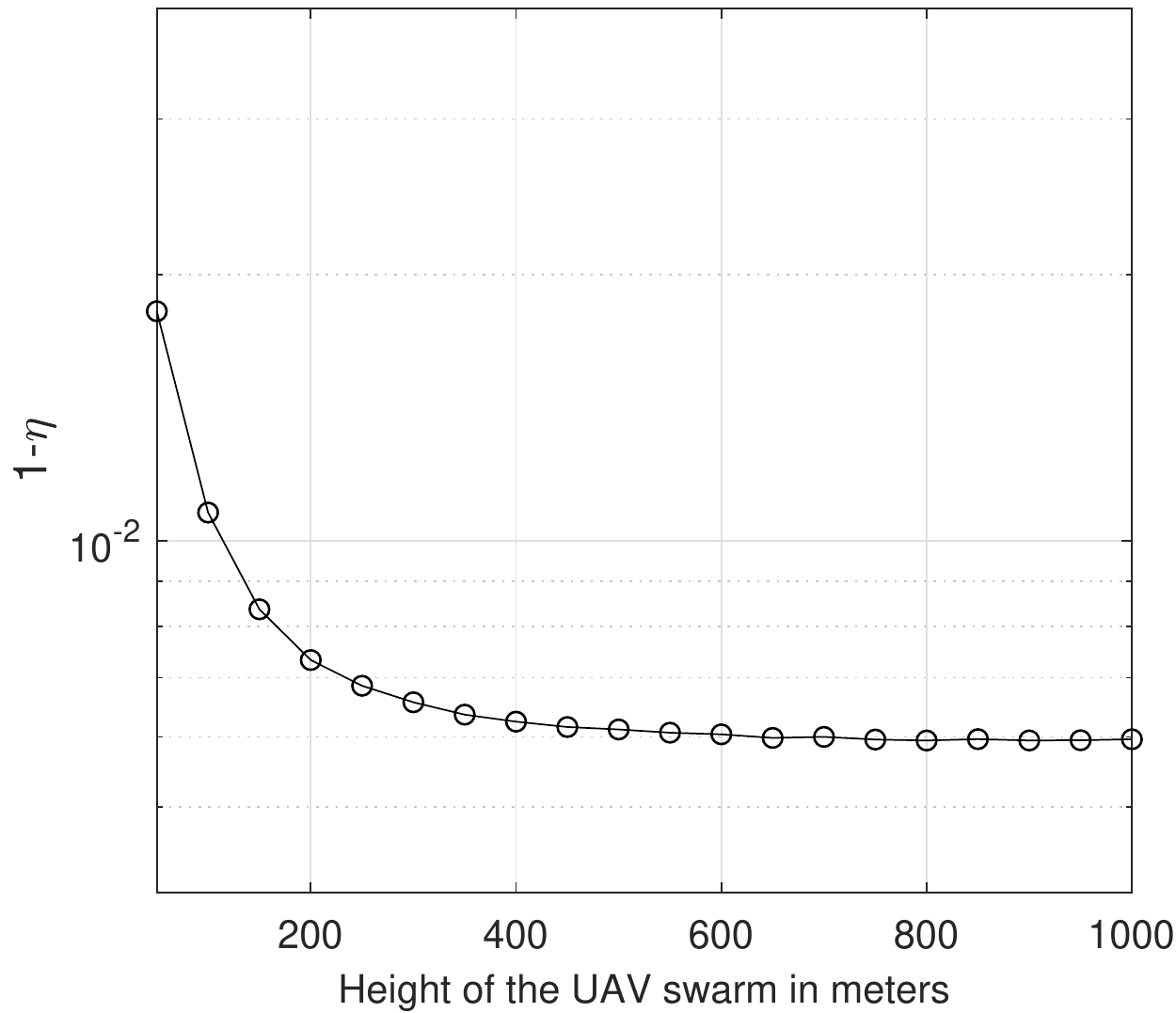}}
\hspace{0.05in}
\subfigure[Effect of the transmission time of Phase I $\tau^{\text{(I)}}$.]{
\includegraphics[width=2.0in]{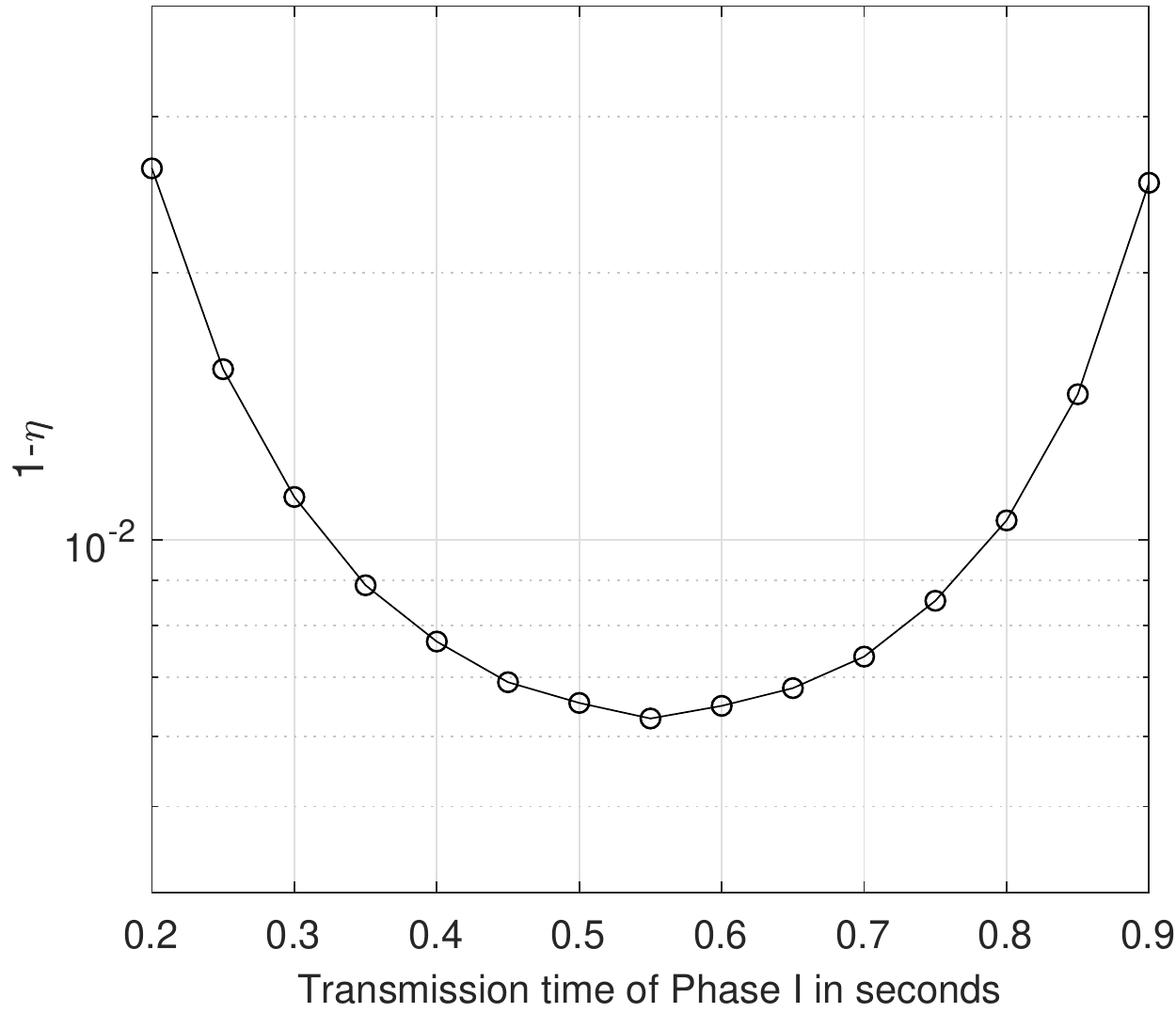}}
\caption{Effect of key system parameters.}
\label{para}
\vspace{-1em}
\end{figure*}

Fig.~\ref{para} (a) shows the effect of the radius of the UAV swarm $\tilde{R}$. As we can see, as the radius of the UAV swarm $\tilde{R}$ increases, $1-\eta$ increases and thus the reliability performance decreases. This is because a larger $\tilde{R}$ gives a larger average distance between UAVs in the swarm, which leads to a higher average path loss in Phase II's D2D communication, in turn a lower reliability performance.

Fig.~\ref{para} (b) shows the effect of the height of the UAV swarm $H$. Note that as the height of the UAV swarm $H$ increases, the swarm should be able to cover a larger ground area, and more available GBSs can help serve it. But letting more available GBSs to serve the swarm can cause a waste of cellular resource, and bring difficulty in synchronizing available GBSs' simultaneous transmission. Thus, here we fix the radius of coverage ground area $R$, as well as the numbers of available and occupied GBSs $(M_0,M_1)$.

As we can see, as the height of the UAV swarm $H$ increases, the reliability performance improves. This is because an increasing $H$ gives a ``smaller'' coverage ground area from the UAV swarm's point of view, i.e., equivalently all the GBSs are ``closer'' to the center of the coverage ground area, therefore the reliability performance increases and finally converges to some value determined by the ratio of the numbers of available and occupied GBSs, i.e., $\frac{M_0}{M_1}$. Note that comparing with the downlink interference generated by occupied GBSs, AWGN is negligible even at $H=1000$ m.

Fig.~\ref{para} (c) shows the effect of the transmission time of Phase I, $\tau^{\text{(I)}}$. As we can see, there exists a trade-off in deciding the transmission time of Phase I, $\tau^{\text{(I)}}$, given the latency requirement $\tau$. A shorter $\tau^{\text{(I)}}$ gives a higher SINR requirement in Phase I, $\theta^{\text{(I)}}$, and on average fewer UAVs can decode the message in Phase I. In this case, although the transmission time of Phase II $\tau^{\text{(II)}} = \tau-\tau^{\text{(I)}}$ is longer, D2D communication still cannot provide a high reliability performance due to its limited power gain. Similarly, for a longer $\tau^{\text{(I)}}$, which gives a lower SINR requirement in Phase I, $\theta^{\text{(I)}}$, but at the expense of a higher SINR requirement in Phase II, $\theta^{\text{(II)}}$. Therefore, we need to balance between two transmission phases for maximizing the reliability performance, and it is optimal to set $\tau_{\text{opt}}^{\text{(I)}}=0.55$ ms given $\tau=1$ ms under this specific setup, as shown in Fig.~\ref{para} (c).

\section{Conclusion}

This paper studies the challenging problem of how to communicate with and control a cellular-connected UAV swarm with both high reliability and low latency. We propose a novel two-phase transmission protocol by exploiting cellular plus D2D communication for the UAV swarm, and characterize the reliability performance of the proposed protocol, i.e., the expected percentage of UAVs in the swarm that can decode the common control message. We show that under reasonable assumptions, it is feasible to decouple the analysis of Phase I and Phase II, and thereby obtain an approximated expression of the reliability performance, with the aid of the Pearson distributions. Numerical results validate the accuracy of our analytical results and show the effectiveness of our protocol over other benchmark protocols. We also study the effect of the swarm head and key system parameters, to reveal useful insights on practical system design.

\section*{Appendix}

\subsection{Proof of Lemma 5.1}

We aim to obtain the approximated distribution of $\overline{X}_{int} = \sum_{m_0\in\mathcal{M}_0}{d}_{m_0}^{\ \frac{-{\alpha}}{2}}|{h}_{1,m_0}^{\text{Rician}}|$. Recall that $M$ GBSs are randomly located inside a circular ground area with radius $R$, following a BPP. Combining with Assumption 4.1, the PDF of the distance ${d}_{m}$ between GBS $m$ and UAV $n$ is approximated by
\begin{equation}
\begin{aligned}
f_{U_{n,m}}(u_{n,m}) = \frac{2u_{n,m}}{{R}^2}, \ \ \ H\leq u_{n,m}\leq \sqrt{{R}^2+H^2}, \ \ \ m\in\mathcal{M},\ n\in\mathcal{N}.
\label{ap1}
\end{aligned}
\end{equation}While the PDF of the small-scale fading $|{h}_{n,m}^{\text{Rician}}|$ between GBS $m$ and UAV $n$ is given by
\begin{equation}
\begin{aligned}
f_{V_{n,m}}(v_{n,m}) \!=\! 2(\kappa\!+\!1)v_{n,m}\exp\big(\!\!-\!\kappa\!-\!(\kappa\!+\!1)v_{n,m}^2\big)I_0\big(2\sqrt{\kappa(\kappa\!+\!1)}v_{n,m}\big), \ v_{n,m}\!\geq\!0,\ m\!\in\!\mathcal{M},\ n\!\in\!\mathcal{N},
\label{ap2}
\end{aligned}
\end{equation}where $I_{0}(z)$ is the modified Bessel function of the first kind with order zero [27, Eq. 8.406.3].

Recall that the path loss and small-scale fading are independent, therefore the mean (first moment) and the variance (second moment) of ${d}_{m_0}^{\ \frac{-{\alpha}}{2}}|{h}_{1,m_0}^{\text{Rician}}|$ can be easily obtained, depending on the path loss exponent ${\alpha}\geq 2$ of ground-to-air channels, i.e.,
\begin{equation}
\begin{aligned}
\mu_{{d}_{m_0}^{\ \frac{-{\alpha}}{2}}|{h}_{1,m_0}^{\text{Rician}}|} & = \int_{H}^{\sqrt{{R}^2+H^2}} u_{1,m_0}^{-\frac{{\alpha}}{2}}f_{U_{1,m_0}}(u_{1,m_0})\,du_{1,m_0}\int_{0}^{\infty}v_{1,m_0}f_{V_{1,m_0}}(v_{1,m_0})\,dv_{1,m_0}\\
& = \left\{
\begin{aligned}
&\frac{\ln(\sqrt{{R}^2+H^2})-\ln(H)}{{R}^2}\times\sqrt{\frac{\pi}{K+1}}L_{0.5}(-\kappa), \ \ \ {\alpha}=4,\\
&\frac{(\sqrt{{R}^2+H^2})^{2-\frac{{\alpha}}{2}}-H^{2-\frac{{\alpha}}{2}}}{{R}^2(2-\frac{{\alpha}}{2})}\times\sqrt{\frac{\pi}{K+1}}L_{0.5}(-\kappa), \ \ \ {\alpha}\neq 4,\\
\end{aligned}
\right.
\label{ap3}
\end{aligned}
\end{equation}where $L_{a}(z)$ is the Laguerre polynomial [27, Eq. 8.970.1], and
\begin{equation}
\begin{aligned}
\nu_{{d}_{m_0}^{\ \frac{-{\alpha}}{2}}|{h}_{1,m_0}^{\text{Rician}}|} \!&= \int_{H}^{\sqrt{{R}^2+H^2}} \!u_{1,m_0}^{-{\alpha}}f_{U_{1,m_0}}(u_{1,m_0})\,du_{1,m_0}\!\int_{0}^{\infty}\!v_{1,m_0}^2f_{V_{1,m_0}}(v_{1,m_0})\,dv_{1,m_0}\!-\!\big(\mu_{{d}_{m_0}^{\ \frac{-{\alpha}}{2}}|{h}_{1,m_0}^{\text{Rician}}|}\big)^2\\
& = \left\{
\begin{aligned}
&\frac{2\big(\ln(\sqrt{{R}^2+H^2})-\ln(H)\big)}{{R}^2}-\big(\mu_{{d}_{m_0}^{\ \frac{-{\alpha}}{2}}|{h}_{1,m_0}^{\text{Rician}}|}\big)^2, \ \ \ {\alpha}=2,\\
&\frac{2\big((\sqrt{{R}^2+H^2})^{2-{\alpha}}-H^{2-{\alpha}}\big)}{{R}^2(2-{\alpha})}-\big(\mu_{d_{m}^{-\frac{{\alpha}}{2}}|h_{1,m}^{\text{Rician}}|}\big)^2, \ \ \ {\alpha}\neq 2.\\
\end{aligned}
\right.
\label{ap4}
\end{aligned}
\end{equation}

Note that the separation distance between GBSs are generally in the order of several hundred meters, therefore from a UAV's point of view, the channels from different GBSs to it can be considered to be independent, more than uncorrelated as explained at the end of Section II. As a result, the elements inside $\overline{X}_{int} = \sum_{m_0\in\mathcal{M}_0}{d}_{m_0}^{\ \frac{-{\alpha}}{2}}|{h}_{1,m_0}^{\text{Rician}}|$ are assumed to be independent and identically distributed (i.i.d.), thus the mean of $\overline{X}_{int}$ is $\mu_{\overline{X}} = M_0\mu_{{d}_{m_0}^{\ \frac{-{\alpha}}{2}}|{h}_{1,m_0}^{\text{Rician}}|}$, and the variance of $\overline{X}_{int}$ is $\nu_{\overline{X}} = M_0\nu_{{d}_{m_0}^{\ \frac{-{\alpha}}{2}}|{h}_{1,m_0}^{\text{Rician}}|}$. $\overline{X}_{int}$ can be approximated by the Pearson type III distribution (the Gamma distribution), with PDF given by
\begin{equation}
\begin{aligned}
f_{\overline{X}_{int}}(\overline{x}_{int}) = \frac{b_{\overline{X}}^{(a_{\overline{X}})}}{\Gamma(a_{\overline{X}})}\overline{x}_{int}^{(a_{\overline{X}}-1)}e^{-b_{\overline{X}}\overline{x}_{int}}, \ \ \ \overline{x}_{int}\geq 0,
\label{ap5}
\end{aligned}
\end{equation}
where $a_{\overline{X}} = \frac{\mu_{\overline{X}}^2}{\nu_{\overline{X}}}$ and $b_{\overline{X}} = \frac{\mu_{\overline{X}}}{\nu_{\overline{X}}}$, which are obtained by matching the mean $\mu_{\overline{X}}$ and the variance $\nu_{\overline{X}}$ of $\overline{X}_{int}$.

\subsection{Proof of Lemma 5.2}

Following similar steps as in \eqref{ap1}--\eqref{ap5}, $\underline{X} = \sum_{m_1\in\mathcal{M}_1}{d}_{m_1}^{-{\alpha}}|{h}_{1,m_1}^{\text{Rician}}|^2$ can be approximated by the Gamma distribution, i.e.,
\begin{equation}
\begin{aligned}
F_{\underline{X}}(\underline{x}) = \frac{\gamma\big(a_{\underline{X}},b_{\underline{X}}\underline{x}\big)}{\Gamma(a_{\underline{X}})}, \ \ \ \underline{x}\geq 0, \nonumber
\end{aligned}
\end{equation}where $a_{\underline{X}} = \frac{\mu_{\underline{X}}^2}{\nu_{\underline{X}}}$, $b_{\underline{X}} = \frac{\mu_{\underline{X}}}{\nu_{\underline{X}}}$, with $\mu_{\underline{X}} = M_1\mu_{{d}_{m_1}^{-{\alpha}}|{h}_{1,m_1}^{\text{Rician}}|^2}$, $\nu_{\underline{X}} = M_1\nu_{d_{m_1}^{-{\alpha}}|{h}_{1,m_1}^{\text{Rician}}|^2}$, and
\begin{equation}
\begin{aligned}
\mu_{{d}_{m_1}^{-{\alpha}}|{h}_{1,m_1}^{\text{Rician}}|^2} & = \int_{H}^{\sqrt{{R}^2+H^2}} u_{1,m_1}^{-{\alpha}}f_{U_{1,m_1}}(u_{1,m_1})\,du_{1,m_1}\int_{0}^{\infty}v_{1,m_1}^2f_{V_{1,m_1}}(v_{1,m_1})\,dv_{1,m_1}\\
& = \left\{
\begin{aligned}
&\frac{2\big(\ln(\sqrt{{R}^2+H^2})-\ln(H)\big)}{{R}^2}, \ \ \ {\alpha}=2,\\
&\frac{2\big((\sqrt{{R}^2+H^2})^{2-{\alpha}}-H^{2-{\alpha}}\big)}{{R}^2(2-{\alpha})}, \ \ \ {\alpha}\neq 2, \nonumber
\end{aligned}
\right.
\end{aligned}
\end{equation}and
\begin{equation}
\begin{aligned}
\nu_{{d}_{m_1}^{-{\alpha}}|{h}_{1,m_1}^{\text{Rician}}|^2} \!&= \int_{H}^{\sqrt{{R}^2+H^2}} \!u_{1,m_1}^{-2{\alpha}}f_{U_{1,m_1}}(u_{1,m_1})\,du_{1,m_1}\!\int_{0}^{\infty}\!v_{1,m_1}^4f_{V_{1,m_1}}(v_{1,m_1})\,dv_{1,m_1}\!-\!\big(\mu_{{d}_{m_1}^{-{\alpha}}|{h}_{1,m_1}^{\text{Rician}}|^2}\big)^2\\
& = \frac{(\sqrt{{R}^2+H^2})^{2-2{\alpha}}-H^{2-2{\alpha}}}{{R}^2(1-{\alpha})}\times\frac{2+4\kappa+\kappa^2}{(\kappa+1)^2}-\big(\mu_{{d}_{m_1}^{-{\alpha}}|{h}_{1,m_1}^{\text{Rician}}|^2}\big)^2. \nonumber
\end{aligned}
\end{equation}

\subsection{Proof of Proposition 5.1}

We aim to solve the following integral
\begin{equation}
\begin{aligned}
\int_{0}^{\infty}\bigg(\int_{0}^{zx}f_{Y}(y)\,dy\bigg)f_{X}(x)\,dx, \nonumber
\end{aligned}
\end{equation}with $F_{X}(x) = \frac{\gamma(c,dx)}{\Gamma(c)},\ x\geq 0$, and $F_{Y}(y) = \frac{\gamma(a,b\sqrt{y})}{\Gamma(a)},\ y\geq 0$,
\begin{equation}
\begin{aligned}
&=\frac{d^c}{\Gamma(c)}\int_{0}^{\infty}\bigg(\frac{\gamma(a,b\sqrt{zx})}{\Gamma(a)}\bigg) x^{c-1}e^{-dx}\,dx\\ \nonumber
&=\frac{d^c}{\Gamma(a)\Gamma(c)}\int_{0}^{\infty}\bigg(\int_{0}^{1}(b\sqrt{zx})^{a}t^{a-1}e^{-b\sqrt{zx}t}\,dt\bigg) x^{c-1}e^{-dx}\,dx\\
&=\frac{2^{-\frac{a}{2}-c+1}(\frac{b^2z}{d})^{\frac{a}{2}}}{\Gamma(a)\Gamma(c)}\int_{0}^{1}t^{a-1}\bigg(\int_{0}^{\infty}x^{a+2c-1}e^{-\frac{x^2}{2}-(\frac{b\sqrt{z}}{\sqrt{2d}}t)x}\,dx\bigg)\,dt\\
&\overset{\text{(a)}}{=}\frac{2^{-\frac{a}{2}-c+1}(\frac{b^2z}{d})^{\frac{a}{2}}}{\Gamma(a)\Gamma(c)}\int_{0}^{1}t^{a-1}\times\frac{\Gamma(a+2c)}{e^{-\frac{b^2z}{8d}t^2}}D_{(-a-2c)}\Big(\frac{b\sqrt{z}}{\sqrt{2d}}t\Big)\,dt\\
&\overset{\text{(b)}}{=}\frac{2^{-a-2c+1}(\frac{b^2z}{d})^{\frac{a}{2}}\Gamma(a+2c)}{\Gamma(a)\Gamma(c)}\int_{0}^{1}t^{a-1}\bigg(\frac{\sqrt{\pi}}{\Gamma(\frac{a}{2}+c+\frac{1}{2})}{_1F_1}\Big(\frac{a}{2}+c;\frac{1}{2};\frac{b^2z}{4d}t^2\Big)\\
&\ \ \ \ \ \ \ \ \ \ \ \ \ \ \ \ \ \ \ \ \ \ \ \ \ \ \ \ \ \ \ \ \ \ \ \ \ \ \ \ \ \ \ \ \ \ \ \ \ \ \ \ \ \ \ \ \ \ \ \ -\frac{\sqrt{\pi}\frac{b\sqrt{z}}{\sqrt{d}}t}{\Gamma(\frac{a}{2}+c)}{_1F_1}\Big(\frac{a}{2}+c+\frac{1}{2};\frac{3}{2};\frac{b^2z}{4d}t^2\Big)\bigg)\,dt\\
&\overset{\text{(c)}}{=}\frac{(\frac{b^2z}{d})^{\frac{a}{2}}}{\Gamma(a)\Gamma(c)}\bigg(\frac{2^{-a-2c}\sqrt{\pi}\Gamma(a+2c)}{\Gamma(\frac{a}{2}+c+\frac{1}{2})}\frac{\Gamma(\frac{a}{2})}{\Gamma(\frac{a}{2}+1)}{_2F_2}\Big(\frac{a}{2}+c,\frac{a}{2};\frac{1}{2},\frac{a}{2}+1;\frac{b^2z}{4d}\Big)\\ \nonumber
& \ \ \ \ \ \ \ \ \ \ \ \ \ \ \ \ \ \ \ \ \ \ \ \ \ \ \ \ -\frac{b\sqrt{z}}{\sqrt{d}}\frac{2^{-a-2c}\sqrt{\pi}\Gamma(a+2c)}{\Gamma(\frac{a}{2}+c)}\frac{\Gamma(\frac{a}{2}+\frac{1}{2})}{\Gamma(\frac{a}{2}+\frac{3}{2})}{_2F_2}\Big(\frac{a}{2}+c+\frac{1}{2},\frac{a}{2}+\frac{1}{2};\frac{3}{2},\frac{a}{2}+\frac{3}{2};\frac{b^2z}{4d}\Big)\bigg)\\
&\overset{\text{(d)}}{=}\frac{(\frac{b^2z}{d})^{\frac{a}{2}}}{\Gamma(a)\Gamma(c)}\bigg(\frac{\Gamma(\frac{a}{2}+c)}{a}{_2F_2}\Big(\frac{a}{2}+c,\frac{a}{2};\frac{1}{2},\frac{a}{2}+1;\frac{b^2z}{4d}\Big)\\ \nonumber
& \ \ \ \ \ \ \ \ \ \ \ \ \ \ \ \ \ \ \ \ \ \ \ \ \ \ \ \ -\frac{b\sqrt{z}}{\sqrt{d}}\frac{\Gamma(\frac{a}{2}+c+\frac{1}{2})}{a+1}{_2F_2}\Big(\frac{a}{2}+c+\frac{1}{2},\frac{a}{2}+\frac{1}{2};\frac{3}{2},\frac{a}{2}+\frac{3}{2};\frac{b^2z}{4d}\Big)\bigg),
\end{aligned}
\end{equation}where $\overset{\text{(a)}}{=}$ comes from the integral representation of parabolic cylinder function [27, Eq. 9.241.2], i.e.,
\begin{equation}
\begin{aligned}
D_{a}(z) = \frac{e^{-\frac{z^2}{4}}}{\Gamma(-a)}\int_{0}^{\infty}e^{-xz-\frac{x^2}{2}}x^{-a-1}\,dx; \nonumber
\end{aligned}
\end{equation}

\noindent $\overset{\text{(b)}}{=}$ comes from the definition of parabolic cylinder function represented by confluent hypergeometric function of the first kind [27, Eq. 9.240], i.e.,
\begin{equation}
\begin{aligned}
D_{a}(z) = 2^{\frac{a}{2}}e^{-\frac{z^2}{4}}\bigg(\frac{\sqrt{\pi}}{\Gamma(\frac{1-a}{2})}{_1F_1}\Big(-\frac{a}{2};\frac{1}{2};\frac{z^2}{2}\Big)-\frac{\sqrt{2\pi}z}{\Gamma(-\frac{a}{2})}{_1F_1}\Big(\frac{1-a}{2};\frac{3}{2};\frac{z^2}{2}\Big)\bigg); \nonumber
\end{aligned}
\end{equation}

\noindent $\overset{\text{(c)}}{=}$ comes from the Euler's integral transform of hypergeometric function
\begin{equation}
\begin{aligned}
{_{A+1}F_{B+1}}\Big[_{b1,\cdots,b_B,d}^{a_1,\cdots,a_A,c};z\Big] = \frac{\Gamma(d)}{\Gamma(c)\Gamma(d-c)}\int_{0}^{1}t^{c-1}(1-t)^{d-c-1}{_{A}F_{B}}\Big[_{b1,\cdots,b_B}^{a_1,\cdots,a_A};zt\Big]\,dt; \nonumber
\end{aligned}
\end{equation}

\noindent and $\overset{\text{(d)}}{=}$ comes from the property of Gamma function $\Gamma(z+1) = z\Gamma(z)$, and the Legendre duplication formula $\Gamma(z)\Gamma(z+\frac{1}{2}) = 2^{1-2z}\sqrt{\pi}\Gamma(2z)$.

\subsection{Proof of Lemma 5.3}

Following similar steps as in \eqref{ap1}--\eqref{ap4}, $\overline{Y}_{int} = \sum_{m_0\in\mathcal{M}_0}{d}_{m_0}^{-{\alpha}}$ can be approximated by the inverse Gamma distribution, i.e.,
\begin{equation}
\begin{aligned}
f_{\overline{Y}_{int}}(\overline{y}_{int}) = \frac{b_{\overline{Y}}^{(a_{\overline{Y}})}}{\Gamma(a_{\overline{Y}})}\overline{y}_{int}^{(-a_{\overline{Y}}-1)}e^{-\frac{b_{\overline{Y}}}{\overline{y}_{int}}}, \ \ \ \overline{y}_{int}\geq 0, \nonumber
\end{aligned}
\end{equation}where $a_{\overline{Y}} = \frac{\mu_{\overline{Y}}^2}{\nu_{\overline{Y}}}+2$, $b_{\overline{Y}} = (\frac{\mu_{\overline{Y}}^2}{\nu_{\overline{Y}}}+1)\mu_{\overline{Y}}$, with $\mu_{\overline{Y}} = M_0\mu_{{d}_{m_0}^{-{\alpha}}}$, $\nu_{\overline{Y}} = M_0\nu_{{d}_{m_0}^{-{\alpha}}}$, and
\begin{equation}
\begin{aligned}
\mu_{{d}_{m_0}^{-{\alpha}}} = \int_{H}^{\sqrt{{R}^2+H^2}} u_{n,m_0}^{-{\alpha}}f_{U_{n,m_0}}(u_{n,m_0})\,du_{n,m_0}
 = \left\{
\begin{aligned}
&\frac{2\big(\ln(\sqrt{{R}^2+H^2})-\ln(H)\big)}{{R}^2}, \ \ \ {\alpha}=2,\\
&\frac{2\big((\sqrt{{R}^2+H^2})^{2-{\alpha}}-H^{2-{\alpha}}\big)}{{R}^2(2-{\alpha})}, \ \ \ {\alpha}\neq 2, \nonumber
\end{aligned}
\right.
\end{aligned}
\end{equation}and
\begin{equation}
\begin{aligned}
\nu_{{d}_{m_0}^{-{\alpha}}} \!=\! \int_{H}^{\sqrt{{R}^2+H^2}} \!u_{n,m_0}^{-2{\alpha}}f_{U_{n,m_0}}(u_{n,m_0})\,du_{n,m_0}-\big(\mu_{{d}_{m_0}^{-{\alpha}}}\big)^2
 \!=\! \frac{(\sqrt{{R}^2+H^2})^{2-2{\alpha}}\!-\!H^{2-2{\alpha}}}{{R}^2(1-{\alpha})}-\big(\mu_{{d}_{m_0}^{-{\alpha}}}\big)^2. \nonumber
\end{aligned}
\end{equation}

\subsection{Proof of Lemma 6.1}

Similar to Appendix D, we aim to obtain the approximated distribution of $Z_{int} = \sum_{k\in\mathbf{\Theta}^{(\text{I})}}\tilde{d}_{n,k}^{-\tilde{\alpha}}$. Note that the distribution of the distance between two random points inside a circle with radius $\tilde{R}$ is
\begin{equation}
\begin{aligned}
f_{W}(w) = \frac{4w}{\pi \tilde{R}^2}\arccos(\frac{w}{2\tilde{R}})-\frac{2w^2}{\pi \tilde{R}^3}\sqrt{1-\frac{w^2}{4\tilde{R}^2}}, \ \ \ 0\leq w\leq 2\tilde{R}.
\label{ap6}
\end{aligned}
\end{equation}Given the minimal separation distance $d_{\min}$ and the simplified model in Fig.~\ref{smp2} (b), the PDF of the distance between receiving UAV $n$ and transmitting UAV $k$, $\tilde{d}_{n,k}$, can be obtained by truncating \eqref{ap6}, i.e.,
\begin{equation}
\begin{aligned}
f_{W_{n,k}}(w_{n,k}) = \frac{f_{W}(w_{n,k})}{\int_{d_{\min}}^{2\tilde{R}}f_{W}(w_{n,k})\,dw_{n,k}}, \ d_{\min}\leq w_{n,k}\leq 2\tilde{R}, \ n\in\mathcal{N}\backslash\mathbf{\Theta}^{(\text{I})}, \ k\in\mathbf{\Theta}^{(\text{I})}. \nonumber
\end{aligned}
\end{equation}Thus, the mean and the variance of $\tilde{d}_{n,k}^{-\tilde{\alpha}}$ can be obtained by solving the following integrals:
\begin{equation}
\begin{aligned}
\mu_{\tilde{d}_{n,k}^{-\tilde{\alpha}}} &= \int_{d_{\min}}^{2\tilde{R}} w_{n,k}^{-\tilde{\alpha}}f_{W_{n,k}}(w_{n,k})\,dw_{n,k},
\label{ap7}
\end{aligned}
\end{equation}and
\begin{equation}
\begin{aligned}
\nu_{\tilde{d}_{n,k}^{-\tilde{\alpha}}} &= \int_{d_{\min}}^{2\tilde{R}} w_{n,k}^{-2\tilde{\alpha}}f_{W_{n,k}}(w_{n,k})\,dw_{n,k}-(\mu_{\tilde{d}_{n,k}^{-\tilde{\alpha}}})^2.
\label{ap8}
\end{aligned}
\end{equation}The complete expressions of \eqref{ap7} and \eqref{ap8} are too complicated, but they are constant and can be easily solved by mathematical softwares, e.g., MATLAB.

Given the independency of transmitting UAVs' locations in \eqref{iv29}, the elements inside $Z_{int} = \sum_{k\in\mathbf{\Theta}^{(\text{I})}}\tilde{d}_{n,k}^{-\tilde{\alpha}}$ are assumed to be i.i.d., thus the mean of $Z_{int}$ is $\mu_{Z_{int}} = K\mu_{\tilde{d}_{n,k}^{-\tilde{\alpha}}}$, and the variance of $Z_{int}$ is $\nu_{Z_{int}} = K\nu_{\tilde{d}_{n,k}^{-\tilde{\alpha}}}$. Let $a_{Z}(K) = \frac{\mu_{Z}^2}{\nu_{Z}}+2$ and $b_{Z}(K) = (\frac{\mu_{Z}^2}{\nu_{Z}}+1)\mu_{Z}$, and the PDF of $Z_{int}$ can be approximated by
\begin{equation}
\begin{aligned}
f_{Z_{int}}(z_{int})& = \frac{b_{Z}(K)^{a_{Z}(K)}}{\Gamma(a_{Z}(K))}z_{int}^{(-a_{Z}(K)-1)}e^{-\frac{b_{Z}(K)}{z_{int}}}, \ \ \ z_{int}\geq 0. \nonumber
\end{aligned}
\end{equation}

\end{document}